# Charting trajectories of human thought using large language models

Matthew M Nour[1,2,+], Daniel C McNamee[3,*], Isaac Fradkin[4,*], Raymond J Dolan[2,5]

1. Department of Psychiatry, University of Oxford, Oxford, UK
2. Max Planck UCL Centre for Computational Psychiatry and Ageing, University College London, London, UK
3. Champalimaud Centre for the Unknown, Lisbon, Portugal.
4. Psychology Department at the Hebrew University of Jerusalem, Israel
5. Wellcome Centre for Human Neuroimaging, University College London, London, UK

+ Corresponding author. matthew.nour@psych.ox.ac.uk
* Equal contribution

# Abstract

Language provides the most revealing window into the ways humans structure conceptual knowledge within cognitive maps. Harnessing this information has been difficult, given the challenge of reliably mapping words to mental concepts. Artificial Intelligence large language models (LLMs) now offer unprecedented opportunities to revisit this challenge. LLMs represent words and phrases as high-dimensional numerical vectors that encode vast semantic knowledge. To harness this potential for cognitive science, we introduce VECTOR, a computational framework that aligns LLM representations with human cognitive map organisation. VECTOR casts a participant's verbal reports as a geometric trajectory through a cognitive map representation, revealing how thoughts flow from one idea to the next. Applying VECTOR to narratives generated by 1,100 participants, we show these trajectories have cognitively meaningful properties that predict paralinguistic behaviour (response times) and real-world communication patterns. We suggest our approach opens new avenues for understanding how humans dynamically organise and navigate conceptual knowledge in naturalistic settings.

> *"Since the time of Aristotle,*
> *thinking has been viewed as a temporal sequence of mental events."*
> Ericsson KA and Simon HA. *Protocol Analysis: Verbal Reports as Data* (1993)

> *"Language is the dress of thought"*
> Samuel Johnson, *Lives of the English poets* (1779–81)

Humans possess a remarkable ability to organise knowledge to suit the task at hand, constructing structured internal representations, known as *cognitive maps*, that support inference, prediction and reasoning [1–10]. Language offers our richest window onto cognitive map organisation, capable of encoding rich conceptual content in dynamic and ecologically valid settings [11,12]. Pioneers of the Cognitive Revolution sought to harness this potential through language-based tasks (e.g., "think aloud" protocols). Yet this effort faced formidable challenges in reliably mapping words to mental content [13], leading to a focus on more structured laboratory tasks that are ill-suited to studying thought processes in more naturalistic settings [14,15].

Artificial Intelligence (AI) large language models (LLMs), by representing words as points in high-dimensional vector spaces, offer new opportunities to revisit this challenge. LLM vectorial representations encode vast semantic and syntactic knowledge [16–19] that explains variance in human behaviour and language-evoked neural responses, thus constituting powerful *foundation models* of human cognitive map organisation [20–28]. Notwithstanding this potential, these representations suffer a critical limitation for cognitive science applications, lacking access to the implicit (non-verbalized) contextualising information that guides cognitive map organisation in humans [6,19]. To harness LLMs for cognitive science, there is a need to develop targeted transformations that filter task-relevant information from LLM representations while incorporating implicit contextualising information [19,29,30].

We introduce *VECTOR* (Vector Embeddings, Concept Decoding, and Trajectory ORganisation), a computational framework that aligns LLM representations with human cognitive maps, drawing on neural decoding and AI model distillation techniques [31–33]. We applied VECTOR to narratives from 1,100 participants prompted to describe the Cinderella story and typical daily routines, task framings that condition participants to structure conceptual knowledge in a particular way. By taking account of this implicit contextualising information, VECTOR embeds narratives as geometric trajectories through a transformed representational space, that we refer to as a *schema space,* which more closely approximates the cognitive maps participants exploit in a given task.

We show that schema space trajectories manifest geometric properties indicative of navigation through a meaningful cognitive map representation. These include strong *geometric alignment* and *directional momentum* across participants, as well as discontinuous *jumps*, echoing William James' characterization of thought dynamics as a sequence of *"flights and perchings"* [34]. At a deeper level, schema spaces surface *abstracted temporal structure* that is shared across task conditions, a hallmark of efficient cognitive map organisation [1,3,6,7]. Crucially, schema space trajectory properties predict language-independent behavioural markers, from fine-grained response times to trait-like communication patterns relevant to psychiatric assessment. The results demonstrate the power of aligning LLM representations with human cognition, enabling systematic study of conceptual organisation and thought through naturalistic language.

# Results

## VECTOR and the geometry of thought

VECTOR is a computational pipeline that frames participant narratives as geometric trajectories through cognitive maps, revealing thought dynamics. The pipeline involves three steps: segmentation of narratives into utterances, embedding in LLM-derived semantic space (*Vector Embedding*), and a transformation to cognitively-aligned schema spaces (*Concept Decoding*). See Figure 1A.

For *segmentation* we developed an LLM-based pipeline to parse narratives into utterances, sentence-like units each defining a single narrative concept (e.g., *"Cinderella lives with her stepsisters"*) [11,35,36]. This segmentation delineates mental event boundaries, evidenced in a finding that inter-word response times (RTs) slow significantly at utterance boundaries (linear mixed model effect of utterance boundaries on RT = 0.54 [95% confidence intervals 0.52-0.57] and 0.63 [0.61-0.66], Cinderella and Routine, respectively, both p<0.001, Figure 1B, see Methods).

*Vector Embedding* maps each utterance to a domain-general semantic space, using a pretrained LLM semantic embedding model (here, we use OpenAI's *text-embeddings-3*). This yielded a 1536 dimensional (1536D) vector for each utterance, regardless of how many words the utterance contains. Crucially, while these embeddings capture rich semantic and syntactic information, they lack access to implicit contextualising information that governs how concepts are related in a specific task setting [19,29], rendering them deficient as cognitive map proxies. For example, in the Cinderella condition, the utterances *"Cinderella wears rags"* and *"Cinderella wears a ballgown"* exhibit high semantic similarity (cosine similarity ~0.75, where 1 is maximal and 0 orthogonal) despite describing different points on a narrative timeline, while the utterances *"Cinderella wears a ballgown"* and *"A carriage made of food"* show low similarity (~0.25) despite relating to adjacent narrative events (Figure 2B).

A *Concept Decoding* step overcomes this limitation by transforming LLM semantic space representations to condition-specific schema space that is low-dimensional, sparse, and interpretable with respect to task features (here, our terminology reflects a definition of a schemas as abstracted knowledge structures that capture information about how events unfold in a specific context [7,37]). We defined a separate schema space for Cinderella and Routine conditions. For each condition, we trained regularized logistic regression models to map from an utterance's 1536D semantic space vector to probability distribution over schema events (8 and 11 schema events for Cinderella and Routine conditions, respectively). This resulted in a low-dimensional embedding for each utterance, where each dimension reflects the probability that an utterance maps to a numbered schema event (i.e., dimension 1 maps to event 1; see Table S1 for event descriptions, and Methods for details of LLM-guided procedures used for schema event identification and decoder training).

Concept Decoding yields reliable and meaningful schema space embeddings. Schema event decoders exhibited high cross-validated decoding accuracy (>80%, chance level accuracy ~10%; see Figure 1C) and robust generalisation performance to out-of-distribution test sets (Figure S1). Strikingly, inter-utterance RTs scaled with inter-utterance schema space distances (a proxy for cognitive map distance [20,38]), over and above any variance explained by semantic space distance alone (linear mixed model effect of schema space distance on RT = 0.30, [0.24, 0.35] and 0.12, [0.06, 0.17], Cinderella and Routine, respectively, both p<0.001, Figure 1B).

In essence, VECTOR treats each narrative as a sequence of conceptual states embedded in two representational spaces: a condition-invariant semantic space (1536D) and a condition-sensitive schema space (8D or 11D, Figure 1A & Figure S2). Each space constitutes a different hypothesis for how words map to latent cognitive map states (akin to competing observation functions used when modelling behaviour operating on partially observable states [6]). In subsequent analyses, we tested whether schema spaces serve as superior cognitive map proxies by examining trajectory properties.

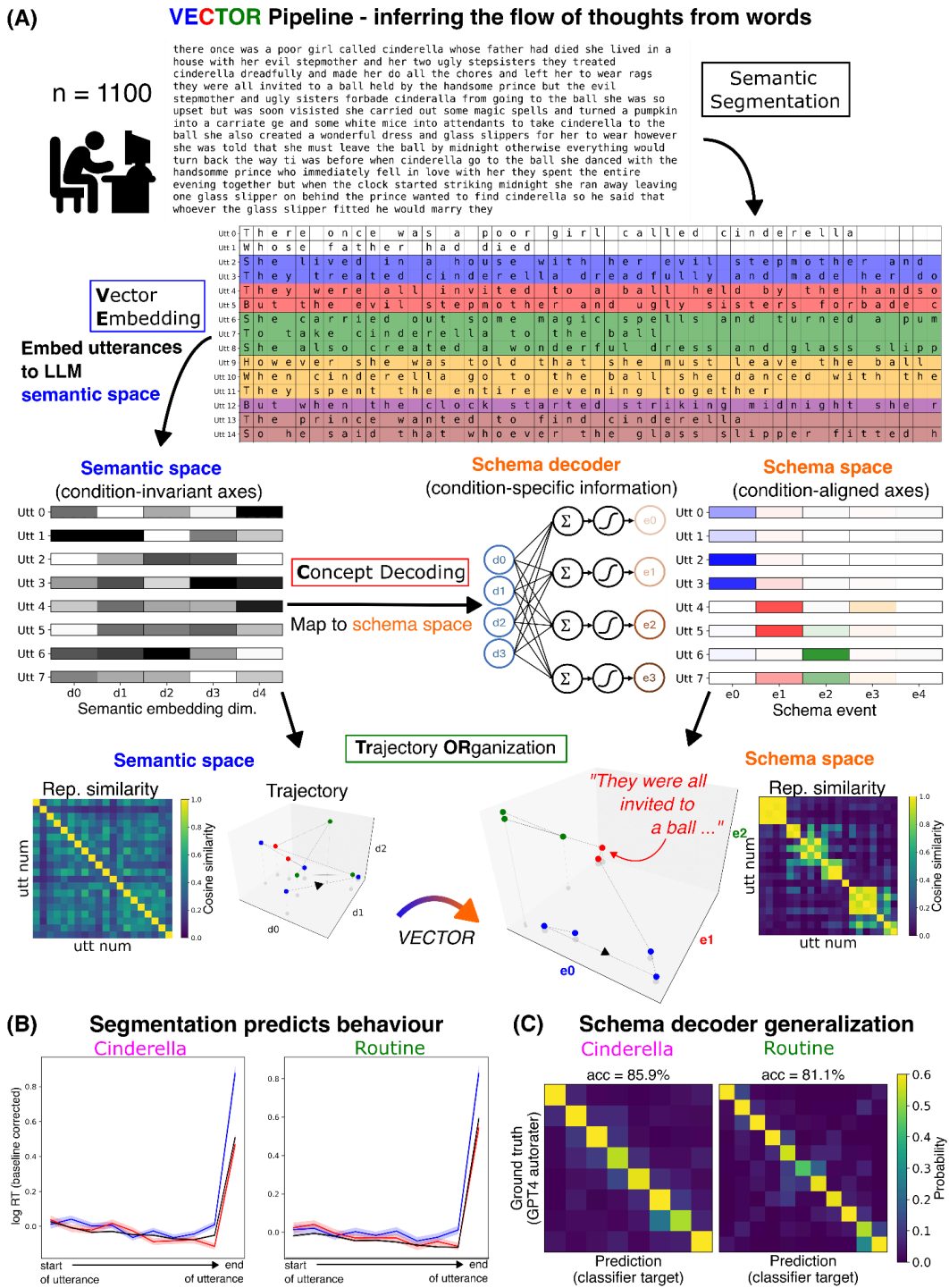

**Figure 1** (legend on next page)

**(Fig 1) VECTOR and the geometry of thought**

**(A)** VECTOR's modular processing pipeline, comprising steps involving Segmentation, Vector Embedding, Concept decoding and Trajectory ORganisation. Illustrated is an exemplar narrative from the Cinderella condition. VECTOR casts a given narrative as a geometric trajectory in two representational spaces - a condition-invariant semantic space and a condition-aligned schema space - where points correspond to utterances. Trajectory plots depict the first 3 dimensions of semantic (left) and schema (right) space. Representation similarity matrices display the cosine similarity between all utterance pairs (ordered by position in narrative) computed using all spatial embedding dimensions, showing event boundaries in schema space alone (block diagonal structure). See Figure S2 & Figure S3 for further illustration, including an exemplar narrative from the Routine condition.

**(B)** VECTOR utterance boundary segmentation and schema space embedding predict participant response times (RTs). Plots show inter-word RTs as a function of progress through VECTOR-defined utterances (mean ± S.E.M. over all narratives), revealing RT slowing at utterance boundaries. RT slowing is more pronounced when the next utterance is far away in schema space (i.e., large conceptual distance). Black lines consider all utterances, while Blue/Red lines consider subsamples of utterances that are followed by large/small jumps in schema space (top/bottom decile of utterance-utterance cosine distances in schema space, respectively).

**(C)** Decoder generalisation performance. Confusion matrices display decoder probability output for utterances of each label class in a balanced held-out test set (all values reflect mean effects over 100 cross-validation folds). See Figure S1 for additional generalisation results. For further computational and data visualization details see Methods.

## Trajectories through cognitive maps

Schema spaces differed markedly from semantic spaces, evidenced both by visual inspection of embedded trajectories (Figure 2A and Figure S3) and a quantitative comparison of representational geometry (Figure 2B & Figure S4).

To test if schema space serves as a meaningful cognitive map proxy we availed of the fact that our experimental design provides ground truth. This is because multiple participants retelling the Cinderella story will necessarily traverse a common sequence of conceptual states. Despite this commonality, the same narrative concept may be described using different linguistic forms, while different concepts may use similar expressions (i.e., state splitting and aliasing [2,6], respectively, see Figure 2B for an example). A valid cognitive map proxy should not be limited by such noise. Rather, it must reliably map (observable) utterances to correct regions of (latent) conceptual space, and structure this space so that representational distance mirrors task structure [1–3,6,7].

We defined two trajectory measures to test these hypotheses. A metric of trajectory *alignment* quantifies the degree to which a given trajectory's transitions can be predicted from patterns across all trajectories (i.e., to what extent does the representation uncover shared trajectories in participants charting similar conceptual paths?). Second, a metric of trajectory *momentum*, quantifies directional progression through space as a function of time [39] (i.e., does representational structure encode linear schematic structure? See Methods for details).

We found that schema space trajectories indeed show significantly greater alignment and momentum than their semantic space counterparts (Figure 2C-D). This signals that VECTOR successfully embeds utterances into a representational space that mirrors cognitive map organisation, recovering shared cognitive organisation between participants, and manifesting a global structure that mirrors the task in question.

We next examined more fine-grained trajectory dynamics. We focused on the intuitive idea that the flow of thoughts is not smooth, but *jumpy,* shifting abruptly from one concept to the next - in William James' words manifesting a series of "*flights and perchings*" [34]. Such jumpiness is expected to manifest as a trajectory characterised by mostly small step sizes, with occasional large jumps [40,41]. While trajectories exhibited significant jumpiness in both schema and semantic spaces, this jumpiness was significantly greater for schema space (Figure 2E, see Methods for details). The fact that schema space jump size also correlated with behavioural slowing ([Figure 1](#)B) suggests that jumpiness captures a genuine property of cognitive map organisation (i.e., discrete latent conceptual attractor states [40,41]), rather than being an artefact of the VECTOR pipeline [20,38].

Finally, as schema spaces have an event-aligned coordinate system - where each dimension reflects an ordered schema event - we asked whether narratives recapitulate the specific linear sequence of conceptual states underlying each task condition. By representing each narrative as an *(event, event)* probability matrix, we recovered an expected signature of forward sequencing. This shows, for example, that a transition from event 1 to event 2 is considerably more probable than from event 2 to event 1 ([Figure 2](#)F).

Overall, our analyses indicate that schema spaces are meaningful cognitive map proxies that reveal an otherwise unseen architecture of thought dynamics. The core intuition here is that LLM semantic embeddings lack access to (non-verblized) contextualising information that scaffolds conceptual organisation, a limitation overcome by VECTOR's Concept Decoding step. We acknowledge that alternative approaches could be employed to transform semantic space embeddings such that they better approximate cognitive maps. In [Figure S5](#) & [Figure S6](#) we outline two candidate approaches - prompt-based contextualization and unsupervised topic modelling [42] - demonstrating a favourable comparative performance to VECTOR's supervised decoding method in reliably uncovering meaningful trajectory signatures. Importantly, VECTOR's use of ground-truth task structure provides a principled benchmark for evaluating these alternative solutions, which are otherwise characterised by sensitivity to experimenter choices in hyperparameter selection [43] (see [Methods](#)).

---

**Figure 2** (next page)
**Trajectories through cognitive maps (Cinderella condition)**
**(A)** Narratives in schema (top) and semantic (bottom) space. From left to right: (i) Utterances embedded in semantic space (all narratives), coloured by each utterance's relative position within the narrative; (ii) Narrative trajectories from 10 participants (points are utterances, coloured by decoded event, trajectories selected on basis of minimal curvature); (iii) Flow fields. The mean trajectory displacement at each spatial location, averaged over all narratives; (iv) Schema events. The mean location of all decoded schema events (centroid) and associated radial histogram, the latter displaying the spatial displacement distribution of trajectory segments starting in the vicinity of each topic centroid.
**(B)** Representational similarity of schema and semantic space. For each narrative, we computed the cosine distance between all utterance pairs in semantic space and schema space, and concatenated this data across narratives. Examples illustrate divergence between cosine similarity in schema and semantic space (see main text).
**(C)** Trajectory alignment. Left: Alignment captures the degree to which trajectories from different participants follow the same path in representational space. Middle: Exemplar trajectories for narratives with highest (black) and lowest (red) alignment scores. Right: Schema space trajectories show higher alignment than their semantic space counterparts.
**(D)** Trajectory momentum. Left: Momentum captures the degree to which a trajectory moves in a directed manner from its starting location [39]. Middle: Exemplar trajectories for narratives with highest (black) and lowest (red) momentum scores. Right: Schema space trajectories show higher momentum than their semantic space counterparts.
**(E)** Trajectory jumpiness. Left: Jumpiness captures the degree to which trajectories manifest small displacements followed by large jumps (i.e., a heavy right tailed distribution of step sizes), quantified by comparing the distribution of observed step sizes to a smoothed trajectory null model [40]. Right: Trajectories exhibited step size distributions that were significantly more heavy tailed than the smoothed null model in both schema and semantic space (95th percentile difference between observed and smoothed step-size distribution: Schema space effect 0.13, p<0.0001; Semantic space effect 0.037, p<0.0001). This effect was greater in schema space vs. semantic space (95th percentile difference 0.092, p<0.0001).
**(F)** Discrete State Sequencing. Left: Sequencing computed on trajectory-level *(event, event)* joint probability matrices by contrasting forward transitions against mirrored backward transitions ("step-1": upper minus lower diagonal mass; "step-n": upper minus lower triangle mass). Middle: Group mean transition probability matrix. Note greater forward vs. backward transition probability. Right: Forward sequencing effects are significantly greater than 0. *Step 1*: Z=490, p<0.001 (One-sample Wilcoxon test); *Step n*: t=68.9, p<0.001 (one-sample t-test).
Throughout, *** indicates p<0.001. Error bars are ± 1 S.E.M. Trajectory visualizations (A, C, D, E) displayed on the first 2 principal axes of each representational space, derived from a space-specific PCA. Alignment and momentum calculated in 2D PCA space; jumpiness calculated 8D (maximal matched dimensionality across spaces). See [Methods](#) for full details of metric computation and data visualisation. All effects replicated in the Routine condition (see [Figure S3](#)).

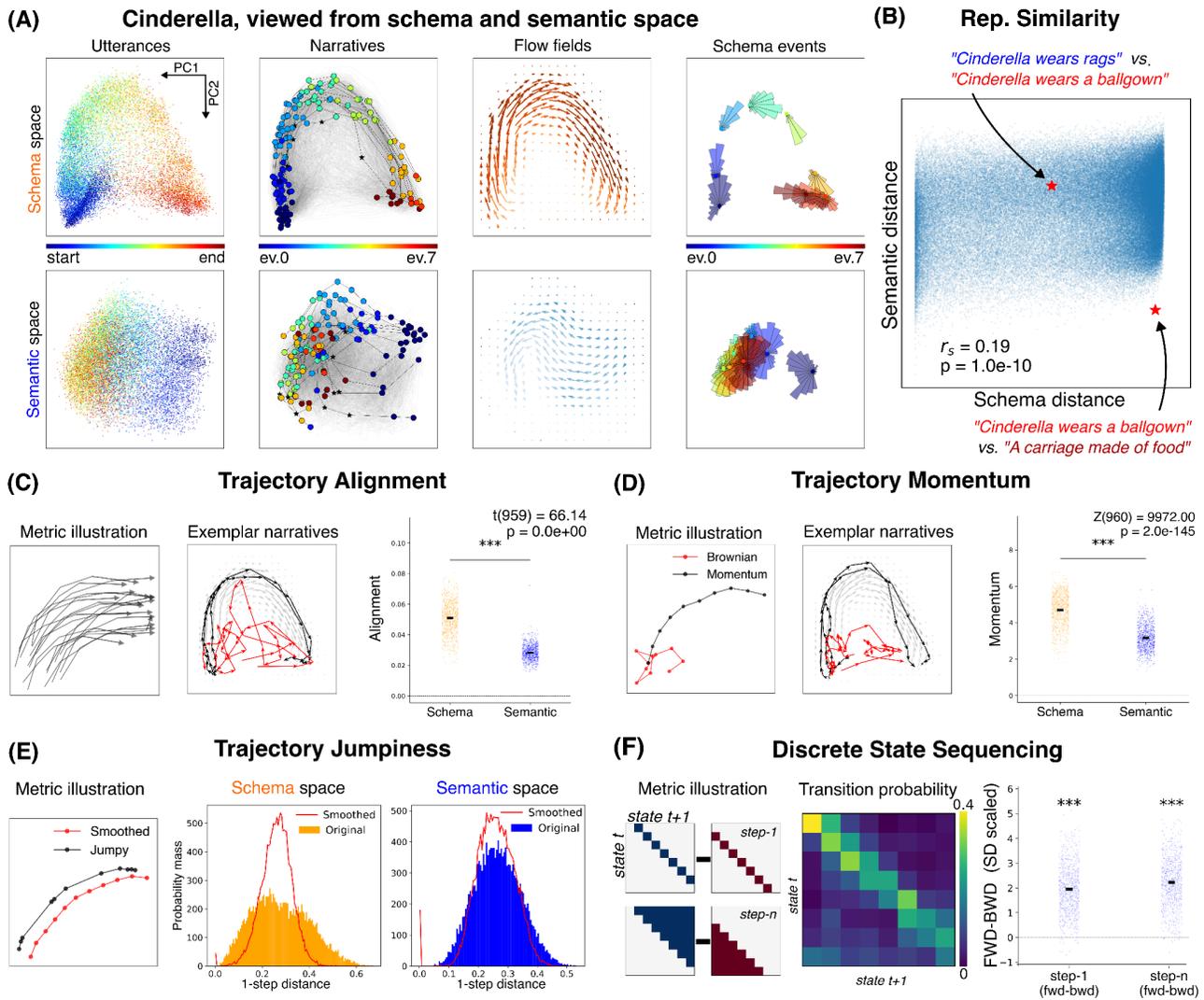

## Transformations for abstraction

The utility of human cognitive maps rests in their ability to organise knowledge as abstracted information, enabling efficient sharing of information across similar contexts [1–3,6,7,10,44,45]. Narrative schemas are a prime example: the event sequences governing one story might differ markedly from another, while still sharing common underlying principles (e.g., that a scene-setting beginning comes before an action-packed middle) [46]. Such reasoning led us to ask whether schema spaces encode abstracted information shared between task conditions.

Taking inspiration from cross-condition generalisation methods in neural decoding [2,45,47], we found that schema decoders trained on one condition (e.g., Cinderella) could uncover meaningful variance in narrative trajectories from an entirely different condition (e.g., Routine). Specifically, when Routine narrative trajectories were projected into Cinderella schema space (and vice versa), we found evidence for forward sequencing (Figure 3A). Strikingly, this generalisation performance extended to an entirely independent narrative generation dataset (n=10,000 GPT3.5/4 generated stories [48], forward sequencing effect highly significant at p<0.001).

This generalisation suggests that decoders, operating on high-dimensional semantic embeddings, have learned patterns that transcend condition-specific content, akin to abstracted temporal information. An intriguing finding from the field of AI mechanistic interpretability is that such abstracted features are often encoded as linear directions in the internal representations of deep neural networks, as one-dimensional projections embedded in high-dimensional embedding spaces [30,49–51]. If such temporal information indeed explains cross-condition generalisation, we hypothesized that it might be possible to identify this temporal feature dimension in semantic space, and use it to steer the output of Cinderella and Routine decoders alike.

We defined a *temporal feature dimension* embedded in high-dimensional semantic space, spanning from start-like concept embeddings (e.g., semantic embeddings of words like *"start"* and *"beginning"*) to end-like concept embeddings (e.g., *"end"*, *"finish"*) [30]. Projecting each narrative utterance onto this vector yields a *temporality score,* capturing the utterance's "start-ness" vs "end-ness" (Figure 3B and Methods). Evidencing the validity of this approach, temporality scores correlated to an utterance's position through its parent narrative (expressed either as relative utterance number [rho=0.25 & 0.43, for Cinderella & Routine, respectively] or decoded schema event number [rho=0.25 & 0.56, for Cinderella and Routine, respectively], all p<0.001, Figure 3C).

In addition to functioning as simple decoding models, feature dimensions can be used as steering vectors, subtly shifting the meaning of a target semantic embedding through vector addition or subtraction [51]. In line with our prediction, schema decoders responded systematically when utterance semantic embeddings were shifted along the temporal feature dimension, with early-event decoders becoming more active when utterances were shifted towards the "start" pole, and late-event decoders becoming more active when shifted towards the "end" pole (by subtracting or adding a scaled feature vector, respectively, Figure 3B-C).

As a final demonstration of latent abstracted temporal information, we used demixed Principal Component Analysis (dPCA) to extract a shared temporal subspace in Cinderella and Routine semantic embeddings directly (Figure 3D) [52]. The first principal component of this temporal subspace was highly meaningful, explaining significant variance in both Cinderella and Routine semantic embeddings (Figure 3E). Strikingly, temporal dPCA also predicted information in an external narrative dataset not used in dPCA model fitting (n=10,000 GPT3.5/4 generated stories [48]). As expected, this effect was not seen when considering the non-temporal (condition-specific) dPCA components, which are insensitive to temporal information shared across narrative conditions (Figure 3F).

Overall, these findings consolidate a view that LLM semantic embeddings contain multi-faceted conceptual information, including abstracted knowledge consistent with a core feature of cognitive map organisation [1–3,6,7]. This information can be uncovered using targeted vector transformations, spanning regression (Concept Decoding), feature vector projection, or variance decomposition.

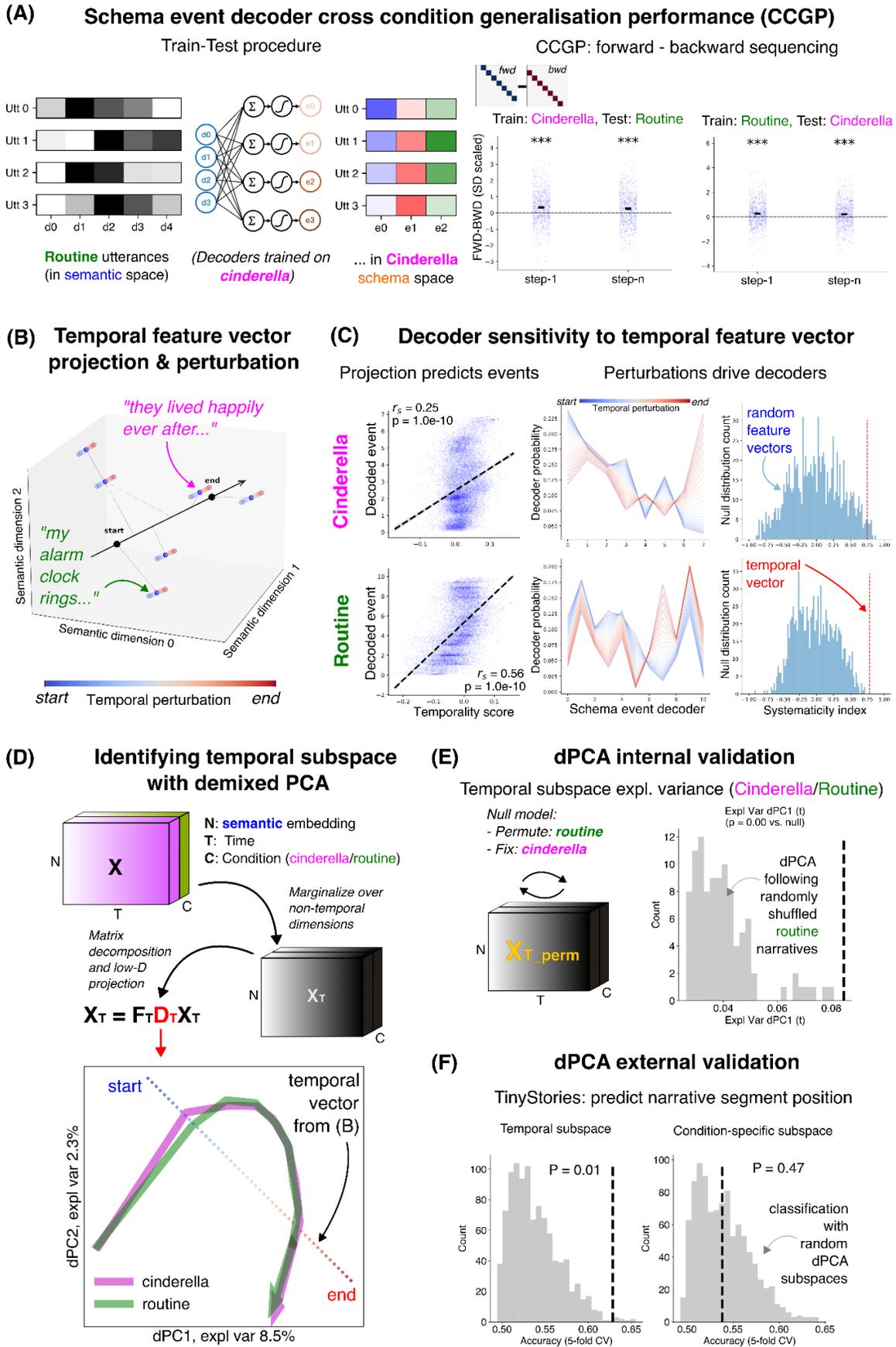

**Figure 3** (legend on next page)

**(Fig 3) Transformations for abstraction**
**(A)** Left: Decoders trained on one condition (e.g., Cinderella) applied to data from the other condition (e.g., Routine). Right: Significant forward sequencing (see Figure 2F) as evidence of cross-condition generalisation. ("train Cinderella, test Routine": *Step 1*: Z=152253, p<0.001; *Step n*: t=8.23, p<0.001. "train Routine, test Cinderella": *Step 1*: Z=161600, p<0.001; *Step n*: Z=172884, p<0.001. One-sample t-test and one-sample Wilcoxon test. *** indicates p<0.001. Error bars are ± 1 S.E.M.)
**(B)** A feature vector in semantic space that captures abstracted temporal information (inspired by [30]). This vector can be used to compute the temporality score of an utterance (vector projection) and modulate its encoded temporal information (vector perturbation).
**(C)** Left: Temporality scores predict an utterance's decoded schema event (spearman correlation, $r_s$; lines represent linear best fit ±95% confidence intervals). Middle: Perturbing utterances along the temporal feature vector predictably drives mean decoder output (early decoders responsive to "start-like" perturbations, late decoders responsive to "end-like" perturbations). Right: Systematicity index, which summarises this response tendency over the complete decoder set, significantly exceeds chance (Cinderella p=0.042, Routine p<0.001, permutation test using random temporal feature vector rotations).
**(D)** Top: Demixed PCA (dPCA [52]) identifies low-dimensional semantic embedding subspaces that capture temporal information shared across conditions. Bottom: mean narrative trajectories projected on the 1st 2 principal components of the temporal dPCA decoder space.
**(E)** Temporal dPCA (1st principal component) captures significant variance in parent narratives (vs. a permutation-derived null distribution that shuffles Routine narratives with respect to Cinderella narratives, breaking any cross-condition temporal coupling).
**(F)** Left: Generalization performance. The temporal dPCA decoder, learned from Cinderella/Routine data, predicts whether a narrative segment belongs to the start vs. end of a story in the TinyStories [48] test dataset (cross-validated accuracy = 62.9%, p=0.008, permutation test using random decoder projections). Right: Generalization not seen when using condition-specific dPCA decoder that is insensitive to abstracted temporal information (accuracy = 53.8%, p=0.44).
See Methods for full details of computational procedures, metric definitions, and permutation testing.

## Trajectories through cognitive maps predict individual differences

In a final, forward looking, validation we asked whether individual differences in trajectory properties are relevant to variance in real-world behaviour. Schema trajectory metrics showed within-participant stability across conditions, suggesting they capture trait-like behavioural patterns (cross-task pearson correlation for alignment r=.09, p=0.005; momentum r=.12, p<0.001; forward sequencing r=.10-15, p≤0.005). Consequently, we examined whether trajectory properties, averaged over conditions, predicted individual differences in self-reported communication atypicality, focussing on a psychometric dimension, termed *eccentricity* [53]. In factor analysis of 128 self-report questionnaire responses, eccentricity loads onto items such as *"I use long and unusual words to say simple things"* and *"People sometimes comment on my unusual mannerisms and habits"* (see [53] and Methods for details).

Individuals with higher eccentricity produced narratives that were more unpredictable, indexed by lower trajectory alignment, momentum, and forward sequencing, as derived from schema space trajectory metrics (Figure 4A). By contrast, semantic space trajectory metrics showed no such association (correlation with alignment rho=-.02, p=0.58; momentum r=-.-05, p=0.11). Moreover, when schema and semantic space metrics were directly compared in multiple regression, schema space trajectory metrics alone predicted eccentricity, consistent with capturing relevant cognitive information not present in untransformed semantic space embeddings (Figure 4B). We confirmed this relationship between eccentricity and trajectory predictability was robust using a complementary LLM-based measure of predictability termed *trajectory entropy*, inspired by work in AI interpretability [54] and the ability of LLMs to simulate individual-specific behaviour [55–57] (Figure 4C). Unlike our alignment measure, trajectory entropy takes account of the entire narrative history when predicting the next utterance (i.e., a non-Markovian measure, see Methods for details).

In a series of control analyses, we found no evidence for a confounding of these associations by participant-level schema decoder performance, narrative topic coverage, or utterance segmentation quality (correlations with eccentricity all r<|0.01|, p>0.05, see Methods for control metric definitions).

Accordingly, all relationships between schema space trajectories and eccentricity remained significant when controlling for these potential confounds in multiple regression analysis. Finally, individuals with higher eccentricity also manifest reduced expression of abstracted temporal structure in emitted narratives (Figure 4D), as quantified using a dPCA-derived *abstraction score* (see Figure 3D & Methods for details).

Combined, our findings indicate that schema spaces capture meaningful behavioural variance that relates to external measures of real-world communication. Individuals with high self-reported communication atypicality (eccentricity) produce narratives that rank lower in predictability and expression of shared abstracted structure (the latter, a hallmark of efficient knowledge generalisation [1–3,10]). This speaks to the validity of schema space metrics as cognitive map proxies, which in turn derives from their ability to accommodate contextualising information about task structure.

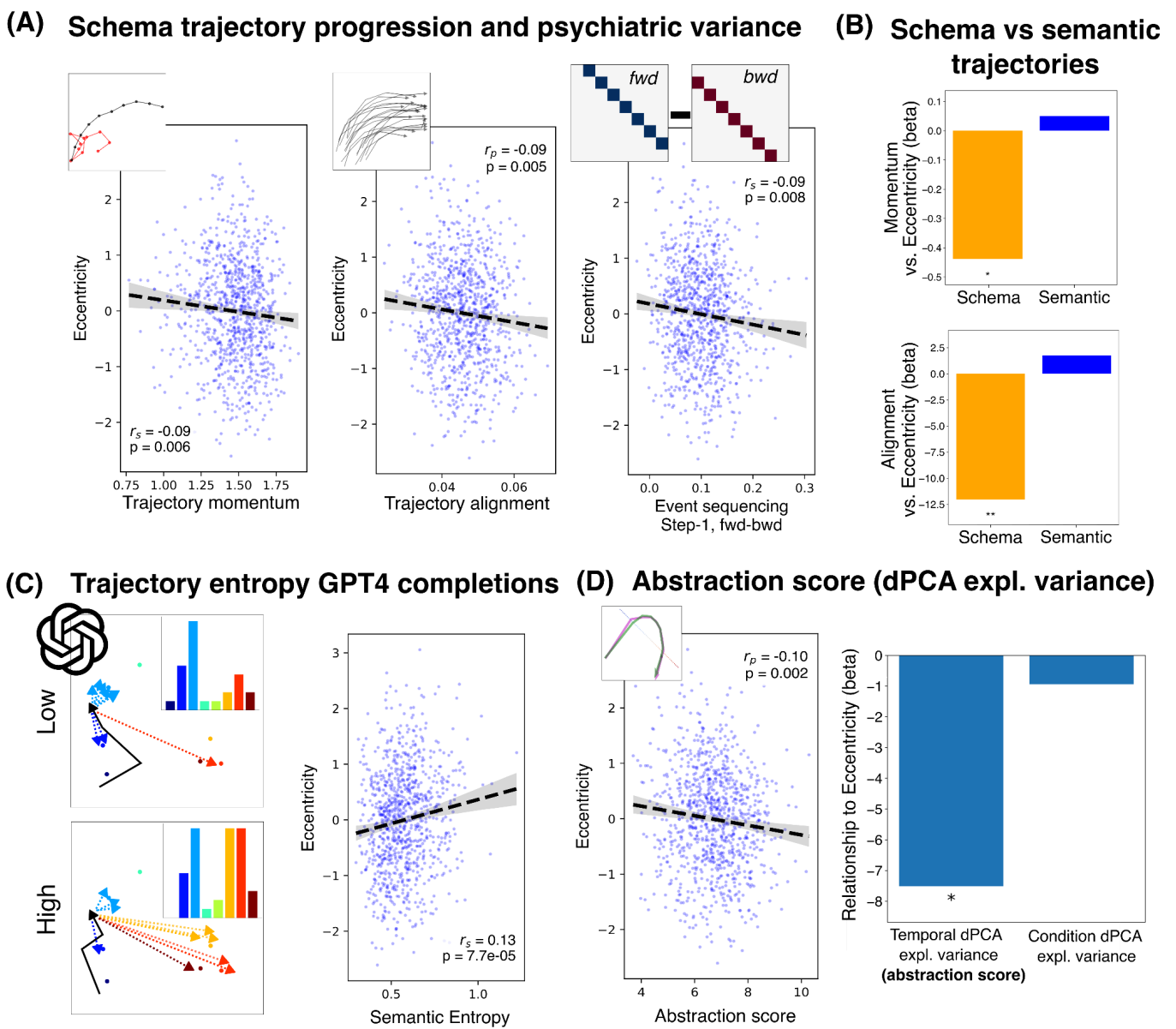

**Figure 4** (legend on next page)

**(Fig 4) Trajectories through cognitive maps predict individual differences**
**(A)** Schema space metrics predict eccentricity scores. From left to right: trajectory momentum, trajectory alignment (both in 2D space, see Figure 2C-D), and discrete state sequencing (step-1 forward sequencing, see Figure 2F).
**(B)** Schema space predictive validity exceeds semantic space predictive validity (schema and semantic space metrics as competing predictors in multiple regression with eccentricity as dependent variable, bar plot shows regression coefficients, β). Top: Momentum, schema space predictors β=-0.44, p=0.03; semantic space predictor β=0.05, p=0.81. Bottom: Alignment, schema space predictors β=-12.0, p=0.01; semantic space predictor β=1.75, p=0.82.
**(C)** Left: Trajectory entropy (illustration), an LLM-based measure of trajectory predictability. For each utterance in a narrative, we used OpenAI's GPT-4o-mini LLM [58] to quantify the entropy of the next utterance's decoded event distribution, conditioned on the entire narrative history up to that point (inspired by [55]). Right: Trajectory entropy (mean of entropy values over all narrative utterances) was positively correlated with eccentricity.
**(D)** Left: Abstraction score (explained variance by temporal dPCA component, Figure 3D) negatively correlates with eccentricity. Right: This relationship remains significant when controlling for explained variance by condition-specific dPCA components in multiple regression (bar plot shows regression coefficients: temporal dPCA predictor β=-7.52, p=0.02; condition dPCA predictor β=-0.94, p=0.47).
All metrics averaged across both conditions per participant. For all scatter plots, lines represent linear best fit ±95% confidence intervals. Pearson correlation ($r_p$) and spearman rank correlation ($r_s$) following the Shapiro-Wilk test for normality. * and ** indicate p<0.05 & p<0.01. See Methods for full details of metric computation.

# Discussion

One of the deepest challenges in cognitive science is understanding how individuals organise and navigate their conceptual knowledge [1–3,6–8,10], a question also central to the study of AI model capabilities [17,59,60]. While language is recognized as our most direct window onto human thought, reliably extracting mental content from speech alone has proven formidable. We provide a framework for solving this challenge, using a computational framework that maps verbal reports to latent concepts embedded in a behaviourally meaningful cognitive map representation. This representation, which we term a schema space, reveals the sequential unfolding of concepts within the stream of thought.

Schema spaces exhibit properties expected of efficient cognitive maps, including low-dimensionality, sparsity, interpretability with respect to task features, and an ability to capture abstracted knowledge that generalises across contexts [1–3,6]. These characteristics contrast with untransformed LLM representations, which are high-dimensional, lack inherent behavioural interpretability, and are relatively insensitive to implicit contextual factors that guide human cognition, spanning implicit situation models and world knowledge [29]. Accordingly, schema space narrative trajectories manifest behaviourally meaningful geometric properties that predict mental event boundaries (as reflected in pauses) and individual differences in communication style.

Core to our approach is the notion that LLM vectorial representations serve as foundation models of human conceptual organisation, encoding knowledge that can be rendered cognitively and behaviourally meaningful through contextualised transformation. This accords with other applications of LLMs in cognitive neuroscience, which employ embedding transformations or model fine-tuning to render LLM representation and behaviour more suited to the question under investigation [26,30,61,62]. It is also supported by principled arguments in AI interpretability, which show that LLM latent representations are best understood as superpositions of decodable features that can be isolated using targeted vector transformations [49,51].

Intriguingly, the challenges addressed by our framework overlap with those faced by the brain when constructing efficient cognitive map representations, a process mandating both (attention-mediated) filtering of irrelevant information and (memory-mediated) introduction of implicit contextualising information [6]. Indeed, theoretical work in cognitive science proposes that task-relevant internal representations are supported by transformation modules acting on domain general representations, where these transformations include concept vector projections [19,29]. At an implementational level, this is thought to necessitate brain systems specialised for encoding domain general relational world models, such as hippocampal-entorhinal circuits, as well as regions that dynamically transform these representations to render them task-relevant, such as medial prefrontal cortex. Here, representational transformations include changes in representational basis and dimensionality reduction [7,37,63–66].

Our framework opens new avenues for psychology and neuroscience, where, to date, the study of learning, inference, and decision making is often constrained to simplified laboratory tasks. These are ill-suited to study the richness and flexibility of cognition in more ecologically valid, open-ended environments [14,15,67]. Accessing cognition using language radically expands this investigational scope. In the specific case of reasoning, we predict that the application of LLM-based methods to verbal report data will renew interest in "think aloud" protocols pioneered during the Cognitive Revolution [13,68].

The approach we detail has considerable translational potential, particularly for understanding cognition in diverse psychiatric conditions, given the scalable and intuitive nature of language-based tasks. Here, we note that VECTOR contrasts sharply with a dominant approach to language analysis in computational psychiatry, which focuses on surface linguistic properties (e.g., word order statistics) or avails of untransformed semantic embedding vectors [69,70]. Arguably the latter misses an opportunity to investigate deeper questions of conceptual organisation and thought dynamics in clinical populations, from rumination to thought disorder, an opportunity now rendered possible by the framework we outline.

We make use of relatively constrained narrative conditions where we also assume simple, known, and conserved latent schematic structure. This facilitates VECTOR's Concept Decoding step and enables comparison to expected ground truth, but also constitutes a significant limitation of the present work. To fully realise the potential of language studies in cognitive science, we will require methodological extensions that address conditions where conceptual structure is poorly understood or more variable.

For instances where task structure is less well understood, we suggest there is value in marrying our task-informed decoding approach with more flexible approaches for identifying latent conceptual structure, including unsupervised topic modelling. As our supplementary analysis shows, however, more unconstrained approaches can be brittle to hyperparameter selection, and thus mandate robust comparisons to external markers of construct or predictive validity.

In the case where we anticipate between-participant variability in cognitive map organisation, our approach enables an adjudication between competing hypotheses. For example, competing hypotheses about cognitive map structure can be encoded as independent Concept Decoding modules or different contextualising text prompts. These can then be ranked according to how well they predict trajectory properties (e.g., trajectory alignment), akin to how one might compare representational motifs in neural data using Representational Similarity Analysis (RSA) [2,44,71,72].

In conclusion, VECTOR enables new opportunities for understanding the architecture of human thought via language, enabling systematic study of cognitive capacities, from creativity to reasoning, that have long resisted quantitative investigation. We anticipate broad applications, spanning cognitive neuroscience through to psychiatry. As we strive to build AI systems that are reliably aligned with human values, developing methods to assess alignment between AI representations and human cognitive structure becomes increasingly critical [73].

# Author contributions

MMN conceived the study hypotheses and analysis, conducted data analysis, and wrote the paper. IF, DM & RJD advised on analysis and edited the paper. IF collected and curated the data and conducted questionnaire factor analysis. RJD secured funding (data collection and OpenAI APIs).

# Funding

This work was supported by an NIHR Clinical Lectureship in Psychiatry to University of Oxford (MMN), a Wellcome Trust Grant for Neuroscience in Mental Health (315364/Z/24/Z) (MMN), an Academy of Medical Sciences Starter Grant (MMN), a BMA Foundation Margaret Temple Award (MMN), a Medical Research Foundation Launchpad Grant (MMN), the European Union's Horizon 2020 research and innovation programme under the Marie Skłodowska-Curie grant agreement (101023441) (IF), and a Wellcome Trust Investigator Award (098362/Z/12/Z) (RJD). The Max Planck UCL Centre is supported by UCL and the Max Planck Society. The Wellcome Centre for Human Neuroimaging is supported by core funding from the Wellcome Trust (203147/Z/16/Z).

# Conflicts of interest

No financial conflicts of interest from any author.

# Supplementary Figures and Tables

| Cinderella | |
|---|---|
| 0 | Cinderella endures a life of servitude under her cruel and wicked stepmother and stepsisters, who subject her to harsh treatment and compel her to do all the household chores. This oppressive environment portrays her as a servant, highlighting the unkindness of her family. |
| 1 | A royal announcement is made for a grand ball at the palace, inviting all eligible maidens in the kingdom, including Cinderella's stepsisters. However, Cinderella wishes to attend but is forbidden by her stepmother, who imposes additional chores on her, leaving her behind while her stepsisters prepare and leave for the ball, which deepens Cinderella's feelings of sadness and isolation. |
| 2 | Cinderella is magically transformed by her fairy godmother, who provides her with a stunning gown, glass slippers, and a carriage made from a pumpkin, allowing her to attend the ball as a beautiful young woman. |
| 3 | Cinderella attends the royal ball, where she enchants and captivates the prince with her beauty and grace, sharing a magical dance that captures his heart; however, she must leave before midnight, when the magic will wear off. |
| 4 | As the clock strikes midnight, Cinderella, in her haste to leave the palace, accidentally leaves behind one of her glass slippers on the stairs or steps. This moment is marked by her urgency as she rushes away, resulting in the loss of the shoe. |
| 5 | The prince embarks on a quest throughout the kingdom to locate the owner of the lost glass slipper, determined to find the woman he danced with at the royal ball. He tries the slipper on every young woman he encounters, ultimately arriving at Cinderella's home in hopes of discovering her identity. |
| 6 | The stepsisters attempt to fit into the slipper but are unsuccessful, revealing their inability to claim the connection to the mysterious girl from the ball. In contrast, when Cinderella tries on the slipper, it fits perfectly, confirming her identity and marking her as the true owner. |
| 7 | Cinderella and the prince are joyfully reunited after he recognizes her as the enchanting girl from the ball, leading to their marriage and her escape from a life of servitude. Together, they embrace a new life filled with happiness, while Cinderella graciously forgives her stepsisters, who face the consequences of their actions. |

| Routine | |
|---|---|
| 0 | The event describes the waking up process of an average person, typically involving the use of an alarm clock to signal the start of the day, followed by the action of turning off the alarm, and often includes getting out of bed in the morning. |
| 1 | The event described involves an average person's morning routine, which includes getting out of bed, brushing their teeth, washing their face, taking a shower or bath for personal hygiene, and getting dressed in suitable clothing for the day's activities. This routine emphasizes the importance of maintaining hygiene and starting the day feeling refreshed and ready. |
| 2 | The event described is the act of preparing and eating breakfast, which is often accompanied by coffee or tea, and is aimed at fueling the body for the day ahead. Breakfast typically includes a variety of foods and beverages to provide the necessary energy for daily activities. |
| 3 | The event described is the daily commute to work or school, which can occur via various modes of transportation such as driving, walking, or using public transport. This routine aspect is consistently highlighted across multiple descriptions, emphasizing the individual's choice of preferred transportation method. |
| 4 | The event describes an average person's routine of attending meetings or classes and engaging in work or study tasks throughout the day, focusing on completing projects, assignments, and responsibilities. This pattern continues into the afternoon, where the individual maintains their efforts until the workday or school day concludes. |

| 5 | The descriptions all refer to taking breaks throughout the day, emphasizing the importance of resting and recharging, whether through relaxation, snacks, or socializing with others. These breaks are typically scheduled or intermittent and are considered essential for maintaining energy levels during the workday. |
|---|---|
| 6 | The event described revolves around having lunch, typically as a midday break from work or studies, which often involves socializing with colleagues or friends. Lunch can be enjoyed at a restaurant or can consist of a packed meal from home, serving as a time to refuel and take a pause from daily responsibilities. |
| 7 | The individual concludes their workday and embarks on their commute home, often taking this time to either reflect on the day's events or unwind from their responsibilities. |
| 8 | The descriptions collectively emphasize the act of preparing and eating dinner, often in a social context, typically involving family or friends, with some variations allowing for the possibility of dining alone or with roommates. The emphasis is on the shared experience of the meal, highlighting the communal aspect of dinner time. |
| 9 | The event described involves an average person concluding their day by engaging in a series of winding down and preparation activities, which include wrapping up daily tasks, planning for the next day, connecting with family or friends, and participating in leisure activities such as exercise, hobbies, or reading. As they transition to bedtime, they follow a nighttime routine that encompasses personal hygiene, skincare, and organizing items for the following day, ultimately reflecting on the day's events and setting goals before going to sleep. |
| 10 | The event described in these various phrases involves the act of going to bed and falling asleep, with the intent of resting and recuperating in order to recharge for the following day, typically lasting between 7 to 9 hours. Setting an alarm is also a common practice to ensure waking up for the next day's activities. |

**Table S1**
**Schema events for each condition**
The composition and order of schema events for Cinderella and Routine conditions was defined using GPT-4o-mini and a consensus clustering procedure (see [Box S1](#)). We used GPT-4o-mini to provide a concise summary of each schema event for use in the LLM-as-judge utterance labelling procedure. We used the following prompt: "*Write a sentence or two summarizing in detail these different descriptions of the same event in {'the Cinderella story', 'the daily routine of an average person'}. Do not describe other events, just the one these descriptions refer to. Do not add further explanatory notes, just stick to your summary:\n\n{list of all the utterances in event cluster, pooled across GPT-4o-mini schema event labelling runs}*", setting the GPT-4o-mini temperature to 1.

**Box S1**

**Identifying condition-specific schema events with GPT4**

To identify the classification targets - the schema structure of each task condition - we prompted GPT-4o-mini to *"List the main events in the story of Cinderella"* or *"List the main events in the daily routine of an average person"*, followed by *"Give your response as a numbered list, with one sentence per event"*. We sampled the model in 10 independent API calls in each condition, yielding multiple candidate events from each API call (each event a sentence or "utterance").

For each condition separately, we then used a consensus clustering procedure to identify a canonical schema event set and ordering. First, we used GPT-4o-mini to generate a "match score" for each pair of utterances generated by the 10 API calls. This score captures the degree to which two utterances belong to the same parent event. For each sentence pair we prompted GPT-4o-mini as follows:

> *{'The story of Cinderella is composed of a series of "events"', 'The daily routine of an average person is composed of a series of "events"'}. For example, here is one possible breakdown of events:*
>
> *{in-context exemplar [74] comprising the ordered utterances from one of the 10 original GPT-4o-mini API calls, that does not include the to-be-rated utterance i or j}*
>
> *I would like you to evaluate whether the following pair of descriptions refer to the same event (return: 1) or different events (return: 0). In the example given above, all possible pairings should have returned 0 (different events). Return only a single number (0 for different, 1 for same). Here are the two descriptions to compare:*
>
> *{utterance i}*
>
> *{utterance j}*

We repeated this procedure 3 times for each utterance pair, on each occasion randomly selecting a different in-context exemplar [74]. We averaged each utterance pair's match scores in an (utterance, utterance) match matrix, and applied the Louvain agglomerative clustering algorithm [75] to identify utterance clusters (candidate schema events). We included a cluster in the final set of schema events if it was represented in at least 7 of the 10 original GPT-4o-mini API calls. The canonical event order was the modal order of appearance of these events (clusters) in the original 10 GPT-4o-mini runs. We used the following default hyperparameters: GPT-4o-mini temperature = 1, match matrix binarization threshold = 0.5, Louvain resolution parameter = 1.

**Box S2**

**Labelling utterances with GPT-4o-mini (LLM-as-judge procedure).**

We used an LLM-as-judge approach 6,7 to generate the supervision labels for classifier training. For each utterance, we prompted GPT-4o-mini to output the single best matching parent event from the canonical schema event set (see Table S1). We used the following prompt:

> *{"The story of Cinderella", "The daily routine of an average person"} consists of the following events*
> 
> *{canonical_events, as per Table S1}*
> 
> *Any utterance taken from a person's {"retelling of the story", "description of a typical daily routine"} can be thought of as belonging to one of these events.*
> 
> *For example:*
> *{a single in-context learning exemplar, containing one LLM-derived utterance for each event, followed by the ground-truth label}*
> 
> *What is the most appropriate event number (0 to {"8", "11"}) that best describes the final utterance of the following passage, taken from one person's retelling of the story. Importantly, while you should take into account the entire passage, you must only give your rating for the final utterance (i.e., what event best describes the place in the story where the passage terminates). Keep in mind that the retelling might not be perfectly accurate and that the events might be out of sequence. Answer '-1' if no event is a good fit. Do not give any further explanation, just a number. The passage to score is:*
> 
> *{target utterances 0:t}*
> 
> *Which event number best describes the last utterance: '{utterance(t)}'?*

When providing a label for utterance $t$, we provided GPT-4o-mini with all preceding utterances in the parent narrative as a context (i.e., utterance $0:t$). We also include an in-context exemplar comprising a high-quality utterance example for each schema event label [74]. Here, each label's exemplar utterance was randomly selected from the GPT-4o-mini utterances that comprised the initial schema event definition step (Box S1), and presented in a random order in the prompt to mitigate prompt sensitivity to ordering effects (e.g., the in-context exemplary might be ordered {1,2,3…} for one API call, but {3,1,2} for another). We applied this LLM-as-judge procedure three times to every utterance of each narrative's best utterance sequence (3 independent GPT-4o-mini API calls), on each occasion using a different in-context exemplar [74]. We defined an utterance's event label as the modal GPT-4o-mini label over these 3 runs (utterances with 3 unique solutions over API calls were not used for decoder training).

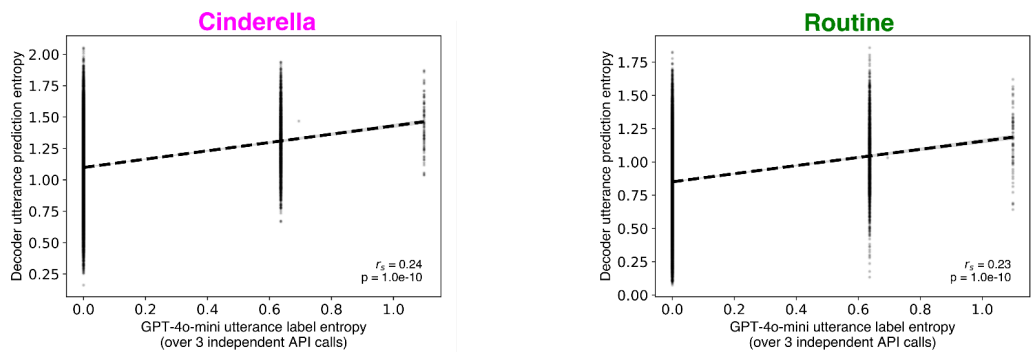

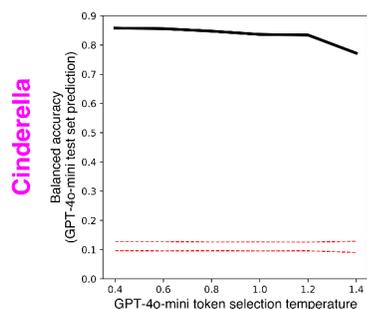

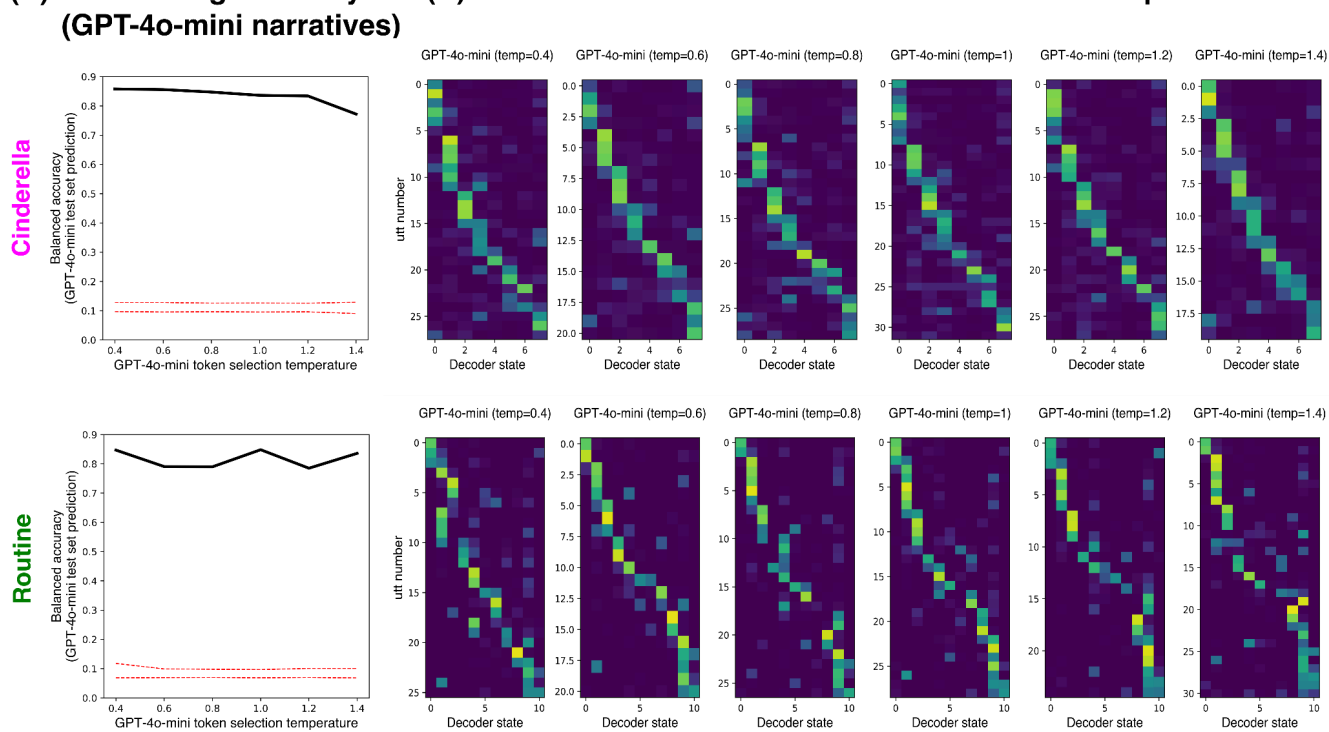

**Figure S1**

**Schema decoder generalisation performance**

**(A)** Schema decoder entropy (the spread of probability mass over schema labels for a given utterance) correlated with the GPT-4o-mini label entropy (the entropy of the 3 schema labels yielded by 3 independent API calls, under the LLM-as-judge procedure, Box S2).

**(B)** Schema decoder generalisation to GPT-4o-mini generated narratives across increasing text generation temperatures. Here, the training data for each condition comprised all labelled participant-derived utterances (n=19,095 and n=16,228, for Cinderella and Routine tasks, form all 1100 participants in the original sample). The test set comprised the concatenated utterances from the GPT4-generated utterances used for the generation of high-quality utterances for VECTOR segmentation, described above) and labelled according to the LLM-as-judge procedure (Box S2). Performance quantified as balanced accuracy. Red dashed lines indicate the 5th and 95th percentiles of the balanced accuracy score from a permuted null distribution of shuffled test-set labels (500 permutations).

**(C)** Decoded GPT-4o-mini narratives across temperatures, with utterances (rows) ordered from start to end of narrative, and events (columns) ordered from start to end of schema event structure (each exemplar randomly selected).

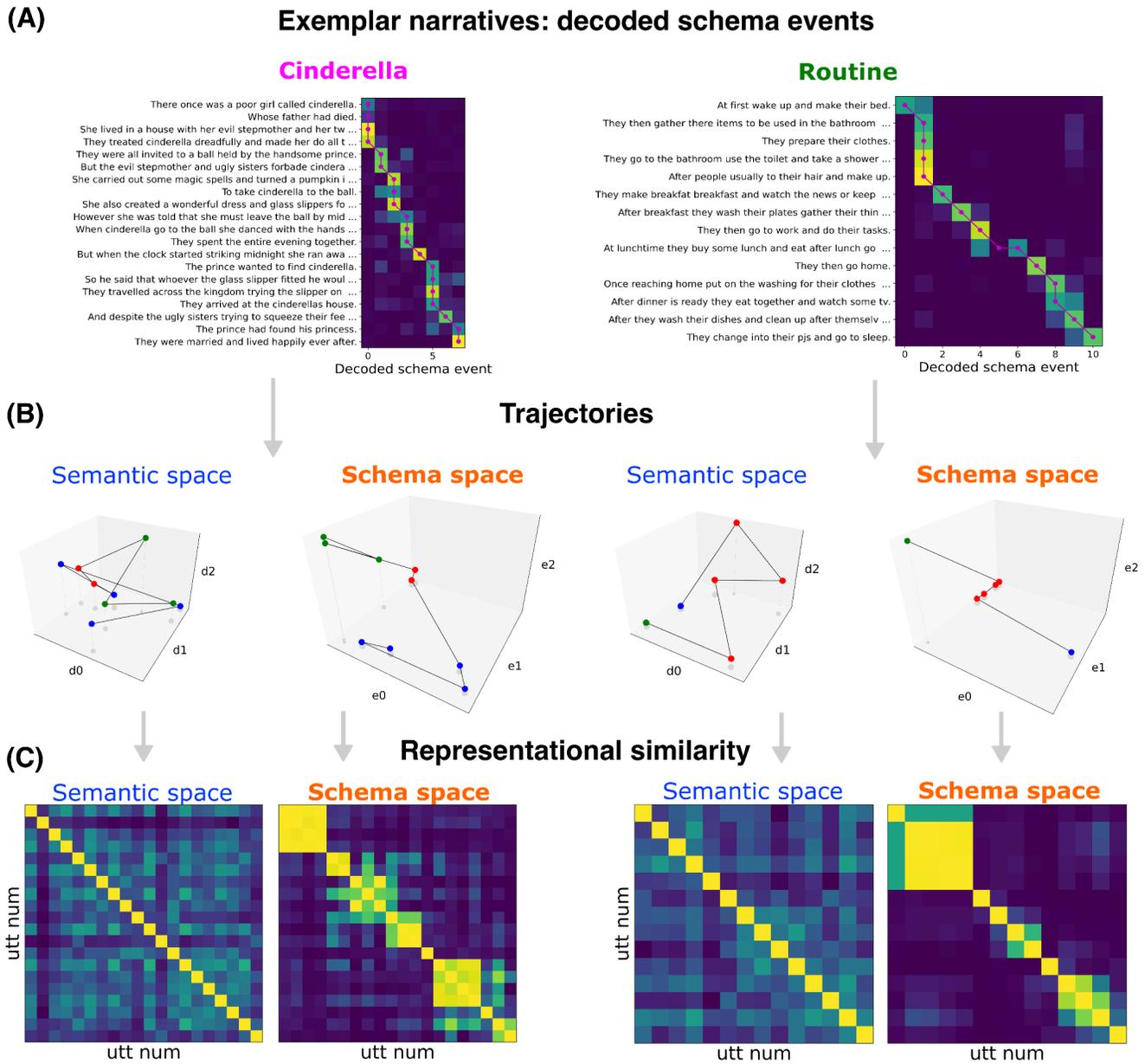

### Figure S2

VECTOR's application to exemplar narrative from Cinderella (left) and Routine (right) condition.

**(A)** Utterance-by-event decoding matrix for each narrative. Schema event decoders are trained in a leave-one-subject out procedure (i.e., the decoders used for both narratives are trained on data from all *other* narratives, and the training set does not include the target narrative)

**(B)** Initial narrative trajectory plotted in 1st 3 axes of semantic (left) and schema (right) space. Utterances (points) colour coded by GPT-4o-mini labelled event number (blue=0, red=1, green=3).

**(C)** Utterance-by-utterance representational similarity, expressed as the cosine similarity between each utterance's semantic (left) or schema (right) embedding vector. Cosine similarity computed using all embedding dimensions. Utterances in the matrix are ordered from start to end (top→bottom; left→right). Colour axes range [0, 1]. Block diagonal structure in schema space matrix is a signature of sensitivity to discrete event boundaries.

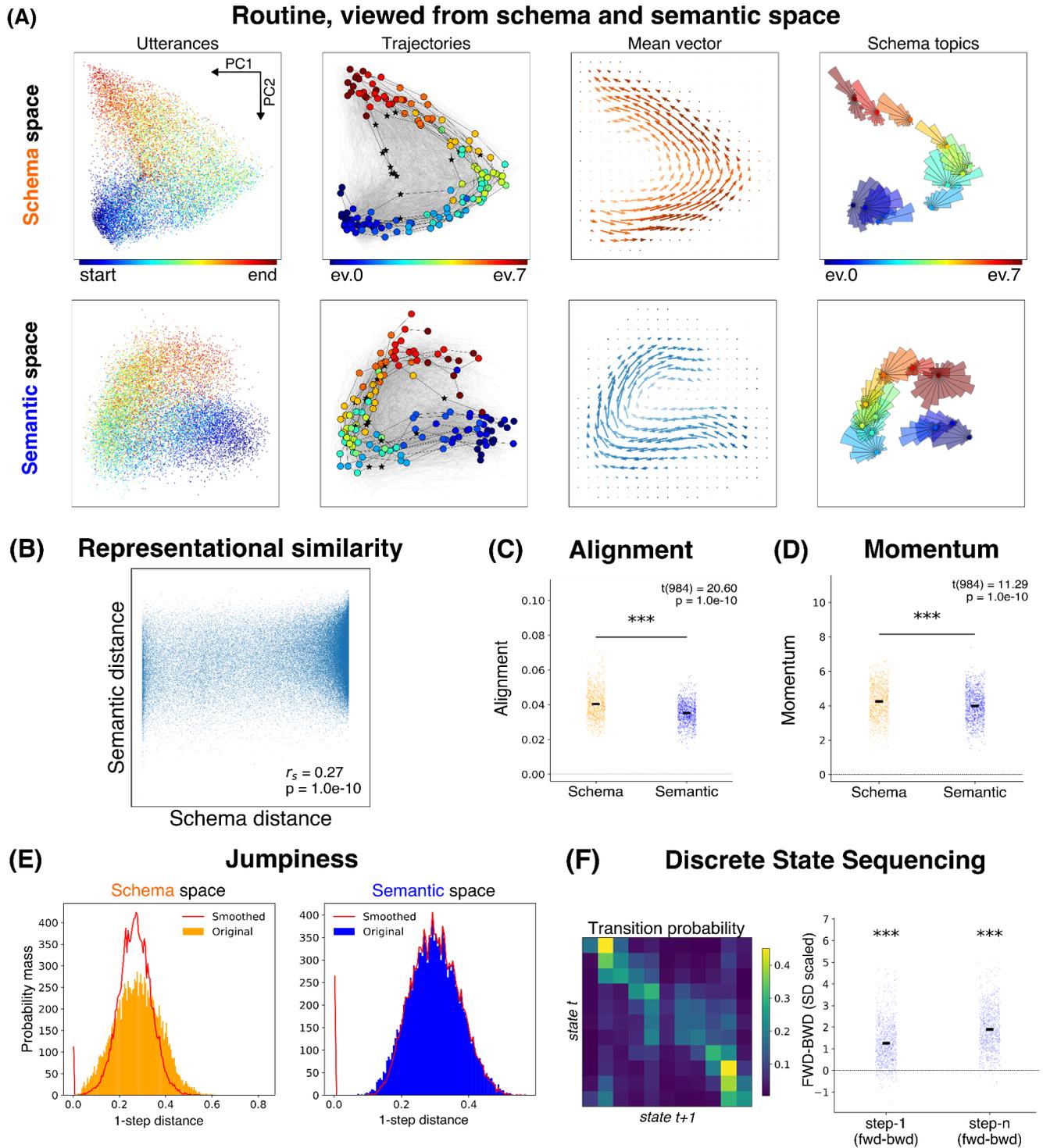

## Figure S3

**Trajectories through cognitive maps (routine condition)**

Visualisation of continuous trajectory metrics for the Routine task condition. Details for panels A-F identical to those outlined for the Cinderella condition visualization (Figure 2). Statistics for **(E)** 95th percentile difference between observed and smoothed step-size distribution: Schema space effect 0.06, p<0.0001; Semantic space effect 0.04, p<0.0001. This effect was greater in schema space vs. semantic space (95th percentile difference 0.013, p<0.0001). Statistics for **(F)** Forward sequencing effects are significantly greater than 0. *Step 1* Z=8122, p<0.001; *Step n* Z=447, p<0.001 (One-sample Wilcoxon test).

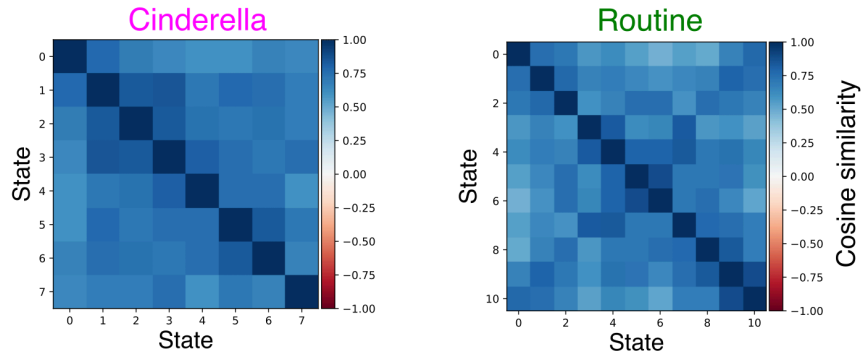

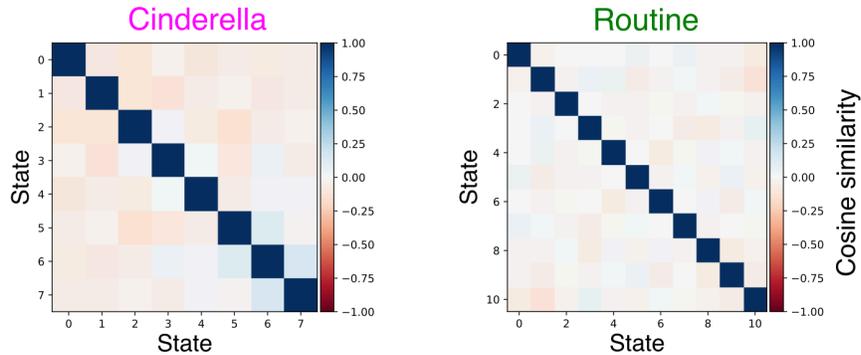

## Figure S4

**Schema decoders disentangle event representations**

**(A)** We computed the cosine similarity between the representation of each pair of schema events in semantic space (encoding weight). This analysis rests on computing a (1536D, number_events) encoding matrix, where each column - a (1536D, 1) vector - corresponds to the typical representation of a given schema event. We computed this matrix by conducting separate multiple linear regressions at each embedding dimension (i.e., 1536 in total), in each case regressing the dimension's (number-utterances, 1) embedding vector on a (number-utterances, number_events) design matrix of GPT-4o-mini dummy-coded utterance labels. Note that the encoding vectors for each event are positively correlated in semantic space (representations of different events lie close together).

**(B)** We contrast this correlated representational encoding structure with the representational structure yielded by schema event decoders, where each events's (1536D, 1) vector corresponds to the weights from training schema event decoders (logistic regression classifiers) to discriminate between different events (as described in [Methods](Methods)). This yielded a more orthogonalised, disentangled representation (near 0 off-diagonal cosine similarities), separating the representation of each event.

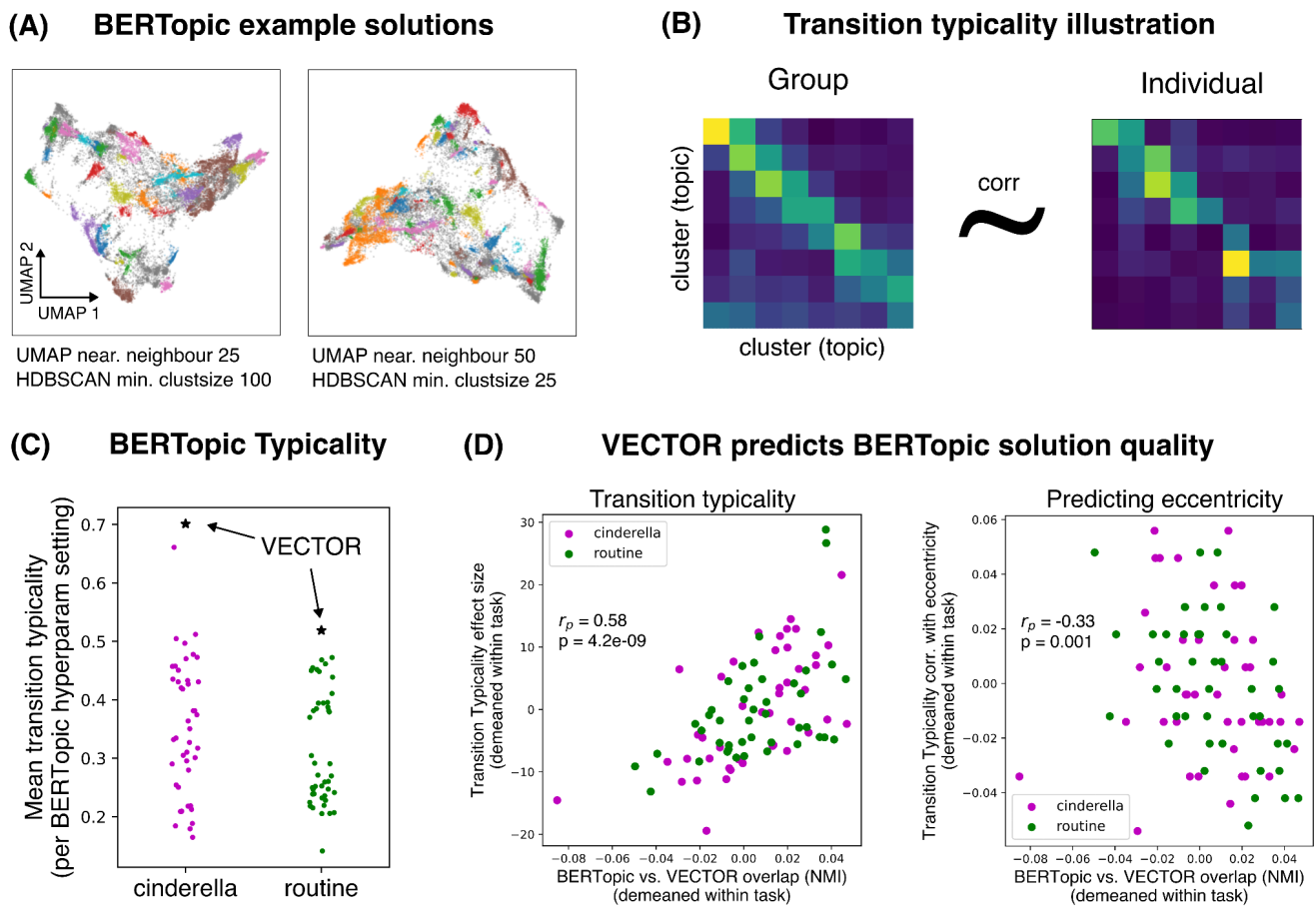

**Figure S5**
**BERTopic: identifying conceptual structure using unsupervised clustering**

**(A)** Two randomly chosen BERTopic solutions of the concatenated Cinderella semantic space utterances, visualized in 2D UMAP space, where each utterance is a point colour coded by HDBSCAN cluster assignment. Note marked sensitivity to embedding geometry and cluster composition (number, size) by hyperparameters.

**(B)** Illustration of transition typicality metric. For a given topic solution (here, illustrated using Cinderella 8 topic [event] schema structure) we compare the (topic, topic) joint probability matrix of each narrative to the group-mean joint probability matrix computed over all other narratives in the group (see [Methods](Methods)). Higher transition typicality scores indicate that a topic solution captures a behaviourally meaningful conceptual structure underlying behaviour.

**(C)** The mean transition typicality scores derived by the BERTopic hyperparameter sweep (n=45 solutions in total) were lower than that observed for the VECTOR-derived topic solution (i.e. 8 schema events for Cinderella; 11 schema events for Routine).

**(D)** BERTopic solution predictive ability increases as the BERTopic solution comes to approximate the VECTOR-derived topic solution. Left: the transition typicality effect (t-statistic) of BERTopic predictive validity (as in (C)). Right: an alternative measure of predictive validity, quantifying the degree to which BERTopic-derived transition typicality negatively correlates with eccentricity. In both plots, x-axis quantifies the degree to which the BERTopic solution overlaps with the VECTOR-based solution (normalised mutual information, NMI), and effects are plotted after mean-centreing within Cinderella/Routine conditions (removing condition-specific baseline effects). Thus, while BERTopic is capable of identifying a latent topic structure that predicts narrative progression and individual differences, the method is highly brittle to hyperparameter choices. Good BERTopic solutions share features with the schema structure identified by VECTOR.

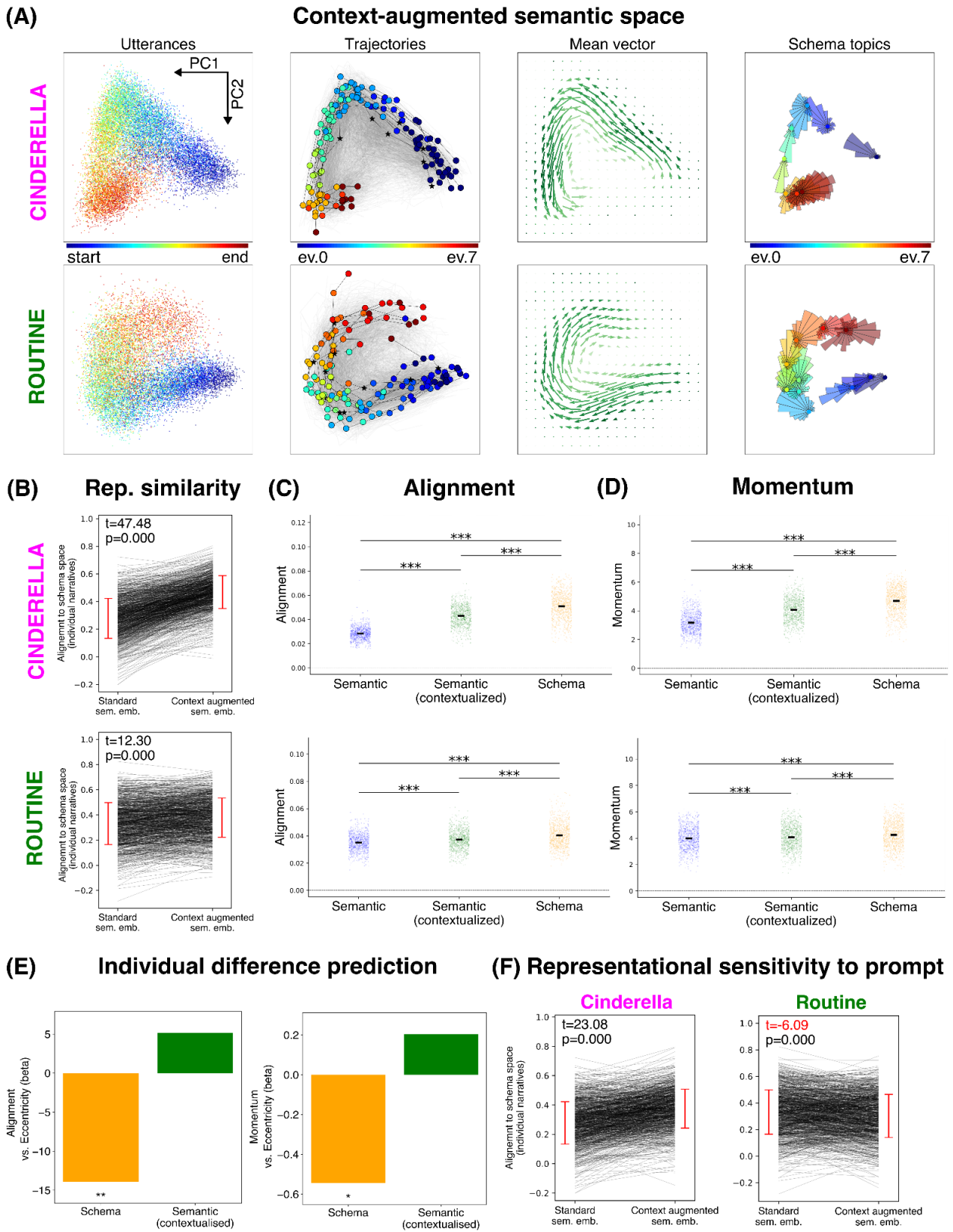

**Figure S6**

**Contextualizing utterances: augmenting semantic embeddings using text prompts**

**(A)** Trajectories through context-augmented semantic space (Cinderella top; Routine bottom). Each utterance was prepended with contextualising text before being embedded in 1536D semantic space (Cinderella: *"In the story of Cinderella: "*; Routine: *"In the daily routine of an average person: "*). Trajectory visualizations then computed identically to [Figure 2](#) & [Figure S3](#). Note visually clear trajectory alignment improvement and topic separation in Cinderella condition in particular (in comparison to non-contextualised semantic embeddings in [Figure 2](#)).

**(B)** Alignment of semantic space representational geometry (non-contextualised standard semantic space, and context-augmented semantic space) and schema space representational geometry. For each trajectory we computed the (utterance, utterance) cosine similarity matrix in 1536D semantic space (both standard and context-augmented) and low-dimensional schema space. We defined cross-space alignment as the spearman correlation between each space's cosine similarity matrix. In both Cinderella (top) and Routine (bottom) conditions, context-augmented semantic embeddings show significantly greater alignment with schema space representational geometry than non-contextualised standard embeddings (paired t-test comparison performed after r-to-z transform).

**(C)** Trajectory alignment scores computed in each space independently. In both Cinderella (top) and Routine (bottom) conditions, while context-augmented semantic embeddings show significantly greater trajectory alignment than non-contextualised standard embeddings, this effect is still lower than trajectory alignment computed with schema space embeddings (*** indicates $p<0.001$ from paired t-test and Wilcoxon signed rank test, following Shapiro Wilk test for normality).

**(D)** As (C), but for trajectory momentum metric.

**(E)** Despite the improved representational geometry of context-augmented semantic embeddings (trajectory alignment and momentum), they still yielded inferior predictive power for individual differences. Plots show regression coefficients for a multiple regression with eccentricity as dependent variable, and trajectory metrics in both schema and context-augmented semantic spaces as competing predictor variables.

**(F)** Illustrative example of the sensitivity of these results to changing the contextualising text. Here, we repeat the analysis in (B), but using a more detailed contextualising prompt for each utterance (Cinderella: *"In the fable of Cinderella, marked by princes, balls and glass slippers: "*; Routine: *"In the daily routine of a typical human, comprising morning rituals, work, and evening wind down: "*). Note the drop in alignment with schema space representational geometry, which, in the case of the Routine condition, drops below that of non-contextualised standard semantic embeddings.

# Methods

## 1. Participants and sample size

The data derive from an online study involving human volunteer participants, approved by the UCL research ethics committee (REC No 16639/001).

1100 participants completed an online language study via Prolific Academic (www.prolific.com). All participants provided written informed consent and were compensated for their time. We used broad inclusion criteria to maximize variation in transdiagnostic symptom variation: age 18-70 years old, raised monolingual with English as a first language, currently residing in the United Kingdom, no colour blindness, minimum approval rate of 95% from at least 50 previous studies, and at most 10,000 submissions on the Prolific platform. Sample demographics are described in [53]: 63% women, mean age 38.2 years (SD 13.01), 60.2% reporting a university degree at bachelors level or higher, ~98% reporting that they received primary and secondary education in an English speaking country.

We used the following exclusion criteria for trajectory analysis and individual differences analyses. First, we excluded participants failing one or more of five attention check questions distributed across the questionnaires (n=100 participants failing at least one attention check) as in our previous study [53]. Additionally, in the Cinderella task (see below), we excluded the 2% of participants who reported not knowing the Cinderella story. Finally, we excluded narratives that included a high proportion of utterances that were deemed to not belong to the task condition in question. Specifically, we excluded narratives where the proportion of utterances assigned "-1" [no event] labels in the LLM-guided labelling procedure (see [Box S2](Box S2) & [Concept decoding](Concept decoding)) exceeded 65% and 71% for Cinderella and Routine conditions, where these thresholds correspond to the mean "-1" assignment + 2.5 standard deviations, over all narratives in each condition.

The final sample was 960 and 985 participant narratives in Cinderella and Routine tasks (used for condition-specific analysis), with 946 participants contributing included narratives in both tasks conditions (used for individual difference analyses, which pooled effects over conditions for each participant).

## 2. Natural Language Tasks

The natural language data acquired from each participant comprised typed text responses. Participants completed two task conditions (order randomized), chosen to promote the use of a latent sequential organizing schema. In the Cinderella condition, participants were asked to "describe in detail the story of Cinderella". In the Routine condition they were asked to "describe in detail what steps are usually involved in people's daily routine". In both conditions, participants were asked to imagine "trying to explain the answer to someone", and provide a narrative account of at least 100 words (2-letter words were not counted towards this minimum narrative length).

To align our typed language protocol with spoken language, the text input interface used in our study ensured that participants could only see the current word on the screen, which disappeared each time they pressed the space or return key to begin the next word. We trained participants on this unconventional mode of typing before the main task, by asking them to copy the task instructions.

## 3. Text preprocessing

We obtained one text sample (narrative) from each participant in each condition. For each narrative, we concatenated individual words into a single whitespace-delimited string. Text was converted to lowercase and stripped of punctuation.

We did not apply processing steps such as stop-word removal or lemmatization for several reasons. First, such preprocessing risks loss of meaningful information. In the present dataset, for example, excluded stop-words account for ~50% of all words [53], and their removal can both distort and degrade semantic content (as an extreme example: *"Cinderella looked **for** a slipper hidden **in** the castle"* and *"Cinderella looked **at** a slipper hidden **by** the castle"*, both become *"Cinderella looked slipper hidden castle"*). Second, this distortion might interact with psychiatric and sociodemographic variables of interest. Finally, stop-word removal and lemmatization might actually degrade performance of applied LLM-based pipelines, given that such preprocessing is typically not applied to LLM training data.

## 4. VECTOR

We designed VECTOR (Vector Embeddings, Concept decoding, and Trajectory Organisation), a sequential processing pipeline, to transform narratives into geometric trajectories through cognitively meaningful representational spaces.

The pipeline involves utterance segmentation, embedding into LLM semantic space (Vector Embedding), transformation into task-aligned schema space (Concept Decoding), and quantification of trajectory metrics (Trajectory Organisation) (Figure 1, note that our utterance parsing procedure itself makes use of semantic space embeddings, hence we discuss these two steps together in the next section).

### Utterance segmentation and vector embedding

The first step segments narratives into utterances, which are sentence-like units capturing single narrative concepts [11,35,36]. Utterances serve as trajectory points.

We developed an automated segmentation pipeline that is sensitive to between-participant differences in expressivity (some participants take more words to communicate single concepts than others), and within-participant differences in expressivity over time (some portions of the narrative progress quickly, with short utterances, while others evolve slowly, taking more words to communicate single details).

We first generate a large set of candidate utterances from each narrative. To do this, we used a BERT LLM ("Bidirectional Encoder Representations from Transformers" [76]) to estimate the probability of an utterance break at each word (operationalized as the union of final softmax probabilities over a stop-token set: [".", "!","?", ";"]). We chose BERT as it is an encoder-only model capable of using text context from both before and after a target [MASK] location to inform its predictions (unlike GPT models, which use an autoregressive next-token-prediction objective and causal masking). This is similar in spirit to the "Where's the Point (WTP)" algorithm, which uses an LLM to predict the probability of a newline token at each character [77].

We performed a separate forward pass of the BERT model for each word $i$ in a narrative. On each occasion we passed the entire narrative as input, with the single addition of introducing a [MASK] token after word $i$ (e.g., $i=0$ *"cinderella* [MASK] *was a housekeeper kept almost like a slave ... "*; $i=1$ *"cinderella was* [MASK] *a housekeeper kept almost like a slave ... "*; etc). We employed a sliding window approach to accommodate narratives that were larger than the BERT model context window (sliding window size = 337 tokens, stride size = 50 tokens; stop probabilities at overlapping [MASK] locations defined as the average over sliding windows).

The above procedure resulted in a vector of "stop probabilities", one for each word location. We used this information to generate a series of candidate utterances for each narrative. Step 1 of this procedure consisted of segmenting the whole narrative into two utterances at the highest ranked break locations (word containing the highest stop probability). Step 2 then segmented each of these two utterances at their respective highest break locations, and so on until we had accounted for the 20% highest break locations. This resulted in a large collection of nested candidate utterances for each narrative (Cinderella: 87,880 unique candidate utterances over all participants; Routine: 74,222 unique candidate utterances over all participants).

To identify the optimal sequence of non-overlapping utterances for each narrative we computed a quality score for each candidate utterance that quantifies the degree to which it captures a single narrative concept. To generate these scores, we embedded each candidate utterance into semantic space using OpenAI's *text-embedding-3-small* embedding model (1536 dimensional (D) vector). We defined a candidate utterance's quality score as the maximal cosine similarity between the utterance's embedding vector, and the embedding vectors corresponding to a large set of LLM-generated condition-specific high-quality utterances (6460 unique utterances for the Cinderella condition and 7251 for the Routine condition). To generate this high-quality training dataset, we prompted GPT-4o-mini [58] to *"Describe in detail the story of Cinderella"* or *"Describe in detail the steps usually involved in a person's daily routine"* (in each case, adding *"Use plain text. Do not use numbered lists or other formatting."*), a total of 50 times for each of 6 model temperature settings between 0.4 and 1.4, inclusive, identifying utterances boundaries by the presence of ".".

In a final step, we used a dynamic programming path finding approach that searched over all candidate utterance combinations to identify the single sequence of utterances in each narrative that maximizes the overall quality score for the narrative. We started by representing each narrative as a directed graph of candidate utterances, where the root node is defined as the start of the narrative ($<START>$, prior to $word_0$), and the leaf node is defined as the end of the narrative ($<END>$, after the last word, $word_N$). Each intervening node, $i$, corresponded to a different candidate utterance. Parents of utterance $i$ corresponded to the set of utterances $j$ in which the final word immediately preceded the first word of utterance $i$ in the narrative. Children of node $i$ corresponded to the set of utterances $k$ where the first word immediately follows the final word of $i$ in the narrative. The $<START>$ node is the parent of all nodes that contain no valid utterance as their parents. The $<END>$ node is the child of all nodes that contain no valid utterance as their child. Each node is associated with a score ($s$, defined above) and a length ($l$ = number words). The algorithm's objective is to identify the path from $<START>$ to $<END>$ that maximises the sum of node-wise scores, with each node's contribution weighted by its length. Utterances with $l < 4$ were not included in the final path, and did not contribute to the scoring (short candidate utterances often contained perseverations and other words that conveyed minimal novel semantic content). Selection of utterances with $l > 25$ was discouraged by artificially setting $s = -1 \cdot 10^5$, thus forcing the algorithm to favor utterances that break large sequences of words into smaller fragments (an idea close to "capped segmentation" approaches taken in recent related work [36]).

We defined each narrative's utterance sequence as the set of candidate utterances comprising the highest-ranking path from $<START>$ to $<END>$. This resulted in narratives with a median of 16 utterances in the Cinderella condition interquartile range [IQR] = 14-19) and 14 utterances in the Routine condition (IQR = 13-16). Utterances comprised a median of 13 (IQR 8-18) words in both conditions. The median total number of words across participants was 212 (IQR 189-259) and 191 (IQR 181-205) for Cinderella and Routine conditions, respectively.

Of note, to ensure the robustness of our subsequent findings to this utterance parsing procedure, we re-ran our subsequent analyses using the 2nd, 3rd, 4th and 5th top-ranking utterance solutions for each narrative, finding equivalent statistical results as when using the single best utterance parsing solution.

## Concept decoding

Concept Decoding implements a targeted transformation on utterance-wise semantic embeddings (1536D, condition-invariant representation), projecting them onto a low-dimensional schema space that is more aligned with cognitive map organisation.

The Concept Decoding approach we use in for the majority of results uses supervised classification algorithms to map each utterance to a probability distribution over condition-specific events (schema events). This approach involved three steps: defining canonical schema events, generating training labels via LLM annotation, and training regularized classifiers.

To define schema events for each condition, we prompted GPT-4o-mini to "*List the main events in the story of Cinderella*" or "*List the main events in the daily routine of an average person*" in 10 independent API calls, and identified a canonical schema event set for each condition using a consensus clustering procedure (see Box S1 for complete details of prompting and clustering). This yielded 8 schema events in the Cinderella condition and 11 in the Routine condition, summarized in Table S1. A canonical event ordering was defined as the modal sequence over all GPT-4o-mini runs containing these events. In this way schema events serve as a latent schematic scaffold for each task condition. A schema space embedding is defined as a vector of weights over these events - 8D or 11D for Cinderella and Routine conditions, respectively.

For each condition, we trained a set of supervised machine learning classifiers to map from semantic space (1536D) to schema space (8D or 11D). We used an LLM-as-judge (auto-labelling) approach [78,79] to generate the supervision labels for classifier training, for each utterance prompting GPT-4o-mini to output the single best matching parent event from the canonical schema event set (see Box S2 for details of this prompting procedure). We repeated this procedure three times for each utterance (independent GPT-4o-mini API calls) and identified each utterance's GPT-4o-mini event label as the modal label over these 3 runs (utterances with 3 unique solutions over API calls were not used for decoder training).

For Concept Decoding proper, we trained a family of one-vs-rest lasso-regularized logistic regression models for each condition: one classifier for each schema event, plus an additional classifier for the "-1" null label. We conducted classifier training across 100 independent cross-validation folds, in each fold defining a balanced test set with one randomly selected utterance per GPT-4o-mini labelled schema event (including the null label, "-1"), and defining a "left in" train set as the concatenated utterances over all participants not contributing to the test set. We used balanced class weighting during training to adjust class weights inversely proportional to training-set class frequencies. Accuracy in each cross-validation fold was defined as the proportion of test utterances where the predicted state (argmax classifier) corresponded to the utterance label (including "-1" label). To identify the optimal regularization parameter for subsequent analysis, we repeated this procedure over a broad set of L1 regularization parameters ($C = 10^k, k \in \{-2, -1, 0, 1, 2\}$, where $C$ is the inverse of regularization strength, such that smaller values indicate greater regularization). Maximal cross-validated generalisation performance was achieved at inverse regularization parameters of 1 and 10 for Cinderella and Routine conditions.

The decoding procedure we implemented accords closely with that adopted in our prior functional neuroimaging decoding studies [45,72]. For all subsequent analyses, we inferred the schema space representation for each utterance using condition-specific decoders, trained using the best-performing hyperparameter value, identified in cross validation ($C$=1 and 10 for Cinderella and Routine, as above).

The final Concept Decoding step applies schema event decoders to each participant's narrative to yield a schema space representation of each utterance (i.e., a probability distribution over schema events), using a leave-one-subject out approach (i.e., decoding on participant $i$'s narrative was performed using models trained using labelled utterances concatenated over all participants $i \neq j$). The schema-space representation used in subsequent trajectory analyses is defined as the vector of decoding probabilities corresponding to each task state classifier (excluding the "-1" null state).

Our approach thus uses information from utterance-wise one-hot labels (derived from GPT-4o-mini) to yield a softer vector representation, which more faithfully captures the global similarity structure between events. This is inspired by the demonstrable power of such soft targets to learn meaningful structured representations in the AI model distillation literature [31,32].

## 5. Measures of trajectory organisation

We defined three metrics to quantify geometric properties of trajectories in semantic and schematic space: alignment, momentum, and jumpiness.

### Dimensionality reduction prior to trajectory metric computation

To allow a direct comparison of trajectories in semantic space (D=1536) and schema space (D=8 or 11), trajectory measures were computed following Principal Component Analysis (PCA) performed in each space independently, with the top $k$ dimensions retained in each space. The input data for PCA was the (utterance, embedding_dimension) data concatenated over all participants, with columns z-scored prior to PCA. Prior to PCA, we applied an exponential temporal smoothing kernel to each narrative's utterance vectors (2-utterance full width half maximum, FWHM; results qualitatively similar when using FWHM=1 or 3 utterances). The resultant principal components were scaled between 0 and 1 prior to trajectory analysis.

We used a dimensionality ($k$) of 2 to compute alignment and momentum trajectory measures. For the jumpiness metric, which was particularly sensitive to the embedding dimension, we report measures that retained as many dimensions as possible while remaining matched between semantic and schema spaces (i.e., $k$=8 for Cinderella, $k$=11 for Routine). Importantly, reported results are robust to the choice of $k$. Specifically, momentum-derived results are robust to choices of $k$ from 2 to 6 (the maximum value examined), and jumpiness-derived results are robust to choices of $k$ from 2 to 8 (Cinderella) or 11 (Routine). Calculation of alignment at higher dimensions is computationally prohibitive as the number of grid points scales exponentially with the number of retained dimensions (see below).

The rationale for fixing this PCA-based step is to compare trajectory metrics calculated at the same dimensionality for both semantic and schema spaces. While the PCA dimensionality reduction described above may appear more drastic for semantic space (e.g., 1536D to 2D), than the schema space (~10D to 2D), recall that both spaces started from the same 1536D representation: the schema space reduction to 2D simply proceeds via two steps (decoding + PCA), in contrast to the semantic space's single PCA step. Given PCA's variance-maximizing objective, of these two procedures, it is actually the semantic space reduction that preserves most of the original variance of the original 1536D embedding.

## Trajectory alignment

Alignment quantifies the extent to which a given participant's narrative trajectory is predictable given a probabilistic model trained on the whole sample. In a procedure inspired by [80], we partitioned semantic and schema spaces into a 20 * 20 square lattice grid extending from the lowest to highest values along each spatial dimension (all reported results are robust to alternative choices of grid resolution: 15*15 and 25*25). Each grid intersection point (20 * 20 = 400) defines the centroid of a spatial state: a circle with radius equal to the grid spacing (i.e., 1/20th the spatial extent). We defined a first-order Markovian transition matrix under this state representation (i.e., a 400*400 Markovian transition matrix), using the trajectory edges from narratives concatenated over all participants. The probability of transitioning between state $i$ and state $j$ is defined as the number of trajectory edges that start in state $i$ and end in state $j$, normalised such that the row-wise probability mass is equal to 1 (or 0). The alignment of a given narrative trajectory was defined as the narrative's mean edge probability, conditioned on this group transition matrix.

## Trajectory momentum

Momentum quantifies the extent to which a given participant's trajectory progresses in a directed fashion through representational space as a function of time. It captures the degree to which trajectory spatial displacement (distance in representational space) correlates with temporal displacement (number of utterances separating two points).

In a procedure inspired by [39], for each trajectory, we fitted a line-of-best-fit relating $\log(t)$ to $\log(x_t^2)$, where $t$ is the number of time-steps (i.e., utterances) separating two trajectory points, and $x_t^2$ is the mean of the squared Euclidean distance separating all utterance embedding vectors $u$ that are separated by a distance of exactly $t$ trajectory edges:

$$x_t^2 = \frac{1}{n} \sum_{i=1}^{n} euclidean(u_i, u_{i+t})^2$$

, where $n$ is the total number of trajectory utterances, minus $t$. Regression slopes >>1 indicate that trajectories display a high degree of directional progression. When fitting lines of best fit we used a maximum temporal displacement ($t$) of 4 (all results are robust to alternative choices of maximal temporal displacements from 3 to 6).

## Trajectory jumpiness

Jumpiness quantifies the extent to which the observed trajectories intermix small transitions in conceptual space with larger jumps, reminiscent of William James' characterization of the stream of consciousness as an unfolding of "flights and perchings" [34], and potentially reflecting the discrete attractor-like properties of underlying representations [40,41]. The null hypothesis is that step sizes sample the trajectory path with equal displacements.

In a procedure inspired by [40], we compared the distribution of observed trajectory spatial displacements to those expected under a smoothed null model. To generate this null model, for each observed trajectory we generated a smoothed mirror trajectory that traversed the same path as the original trajectory but with uniform step sizes. Smoothed trajectories were constrained such that all its points needed to come from the observed dataset of all trajectories, thus ensuring they were composed of points in conceptual space that corresponded to valid utterances (a "snap-to-grid" approach). Repeating this procedure over all trajectories and concatenating the step size vectors yielded two step size distributions for each space: one comprising observed step sizes, and one comprising smoothed step sizes.

Jumpiness manifests as a step size distribution with a fatter right tail than the smoothed distribution (i.e., more large jumps than expected). To test this, we focused on the 95th percentile difference between observed and smoothed step size distributions (i.e., percentile(observed, 95) - percentile(smoothed, 95)), and established statistical significance using permutation tests. For each of 500 permutations, we concatenated observed and smoothed step sizes and randomly reassigned half the elements to "observed" and half to "smoothed" conditions. We then recalculated the 95th percentile difference between (permuted) observed and smoothed distributions, and defined the p-value of this test as the proportion of permuted difference scores that were greater than or equal to unpermuted (actual) difference score (i.e., a right-tailed test). We used the same permutation approach to test whether schema and semantic spaces differed in jumpiness. Here, the effect of interest was the 95th percentile difference between schema and semantic distributions (i.e., percentile(schema, 95) - percentile(semantic, 95)), and the permutation involved shuffling the "schema" vs. "semantic" condition labels on the concatenated step size distribution yielded by both spaces.

In all cases where we report significant differences between step-size distributions using the fat-tail test above, we find similar significant differences between distributions when using the more conventional (but less directionally-specific) Kolmogorov-Smirnov test. The results reported using the fat-tail test are robust to different choices of the percentile hyperparameter (90th & 99th distribution percentile).

### Discrete state sequencing

The above metrics – alignment, momentum, and jumpiness – operate on a continuous space representation. Schema space representations also afford an opportunity to examine how trajectories move through discrete states, as the axes of schema spaces are aligned with interpretable schema events which have an implicit ordering (derived from GPT-4o-mini).

The schema space representation of each narrative can be construed as an (utterance, event) decoding matrix, from which we can construct an (event, event) joint probability matrix, $M$, where $M_i^j$ denotes the probability of occupying state $i$ at time $t$ and state $j$ at time $t+1$, averaged over all consecutive utterance pairs in a narrative.

The tendency of a given narrative to progress sequentially through states manifests as a greater probability mass in the upper vs lower triangles of this joint probability matrix. We captured this using two related metrics. *Step 1* sequencing is defined as the contrast between the 1st off-diagonal of the upper triangle to the 1st off-diagonal of the lower triangle (i.e., comparing "event 1→event 2" to "event 1←event 2" transitions). *Step n* sequencing repeats this procedure using the entire upper/lower triangle mass, relaxing the requirement that transitions need to move to adjacent states while still capturing directionality (i.e., comparing "event 1 → event {2,3,4…}" to "event {1,2,3,…} ← event 1" transitions). In both cases, 0 represents the null hypothesis of no meaningful sequentiality. (Results were robust to using an alternative z-scored metric definition, which quantifies the degree to which an individual trajectory's (event, event) matrix exhibits forward sequencing in a manner that exceeds a trajectory-specific null distribution, where this null distribution is derived by recomputing sequencing measures under 100 event order permutations.)

## 6. Abstracted sequential schema representations

### Cross condition generalisation performance (CCGP)

Cross-condition generalisation performance (CCGP) measures the degree to which schema decoders encode abstracted information that generalises across task conditions, using measures inspired by the neural decoding literature [2,45,47]. To quantify CCGP, we first applied schema event decoders trained on one condition (e.g., n=8 events in Cinderella) to the other condition (Routine), thus projecting each narrative trajectory in the "wrong" schema space.

As our CCGP measure, we calculated sequentiality metrics (see [Discrete state sequencing](#)) for each narrative, projected into a schema space defined by the "wrong" decoder set: Cinderella narratives in Routine space & Routine narratives in Cinderella space. If schema decoders contain no information that generalises across conditions, then the forward-minus-backwards subtraction contrast should yield a distribution of sequentiality metrics centreed on 0.

We also repeated this procedure on an entirely different test dataset: 10,000 stories from the TinyStories dataset, each a short narrative generated by GPT-3.5/4 [48]. To generate trajectories from these stories, we first segmented each TinyStories narrative into 6 segments, and embedded each segment in 1536D semantic space (specifically, we iteratively merged the shortest pairs of consecutive sentences in each narrative to yield a 6-segment solution for each narrative, excluding narratives with <6 sentences).

## Temporal feature vector perturbation analysis

After establishing that CCGP exceeded 0 in the above analysis, we probed deeper into the nature of the abstracted information driving this finding.

Inspired by [30], we constructed a temporal feature vector by computing difference vectors between semantic embeddings of temporally opposing word pairs. We embedded three "start" words (*"first"*, *"beginning"*, *"start"*) and three "end" words (*"last"*, *"end"*, *"final"*) using the same OpenAI *text-embedding-3-small* model used for utterance embedding. We computed all pairwise difference vectors ("end" minus "start" embeddings), yielding 9 temporal feature vectors, which we normalised to unit length and averaged to create a single temporal feature vector.

We validated the meaningfulness of this temporal feature vector as follows. We defined each utterance's *temporality score* as the scalar value resulting from projecting the utterance's 1536D semantic embedding vector onto this temporal feature vector (dot product). We then computed the correlation between utterance-wise temporality scores and both relative position within the parent narratives (utterance number divided by total utterance count) and decoded schema event number (the sum of predicted event indices weighted by each event's decoded probability).

We then performed a perturbation experiment to test schema decoder sensitivity to temporal structure. For each utterance, we displaced the semantic embedding vector by $\alpha \mathbf{v}$, where $\mathbf{v}$ is the unit temporal feature vector, and $\alpha$ is a scalar value uniformly sampled at 25 values from -1 to +1. This resulted in 25 new vector embeddings for the utterance in question, where embeddings corresponding to negative values of $\alpha$ are more "start like" and embeddings corresponding to positive values of $\alpha$ are more "end like". For each new vector we calculated the probability distribution over schema events (excluding the null event label), using pretrained condition specific schema decoders (trained over all participants on labelled, unperturbed data, see [Concept decoding](#)). Repeating this over all utterances yielded a large (utterance_number, perturbation_number, schema_event) array. Computing the mean over the 1st dimension yielded a (perturbation_number, schema_event) array that captures each event decoder's sensitivity to perturbation magnitude ($\alpha$).

The key question is whether an event decoder's sensitivity to perturbation magnitude ($\alpha$) was itself predictable by the event's position in the parent schema (1st event, 2nd event etc). Capturing this information requires us to measure the degree to which an event decoder's responses varied monotonically with $\alpha$. For each event decoder, $i$, we computed the regression coefficient $\beta_i$ that captures the linear relationship between $\alpha$ and decoder output probability. Repeating this over all decoders yielded a (schema_event, 1) vector of $\beta$ values: positive $\beta$ indicates that the event decoder's output increases with higher ("end like") values of $\alpha$, negative $\beta$ indicates that the event decoder's output increases with lower ("start like") values of $\alpha$. Armed with this information, we could ask whether an event decoder's relative sensitivity to "start like" or "end like" perturbations can be predicted from the position of the decoder's event in the parent schema. We operationalised through using a *systematicity index*: a simple Pearson correlation between $\beta_i$ (a decoder's sensitivity to $\alpha$) and $i$ (the ordinal position of the decoder's event in the parent schema).

Statistical significance was assessed using permutation testing with 1,000 random rotations of the temporal feature vector, generated using the special orthogonal group to produce uniformly distributed rotations in high-dimensional space (similar results were obtained using an alternative permutation procedure that fixed the temporal feature vector and instead shuffled perturbation indices).

## Demixed Principal Component Analysis

To explicitly quantify shared temporal structure across Cinderella and Routine narrative embeddings, we applied demixed PCA (dPCA), a matrix decomposition technique that separates variance attributable to different experimental variables [52]. Unlike standard PCA, dPCA learns separate decoder-encoder matrices for different sources of variance. In our case, this involves learning separate matrix decompositions for abstracted temporal information shared across conditions, and information that distinguishes conditions.

As dPCA requires equal-length sequences across conditions, we first binned each narrative into 10 evenly spaced time bins, averaging embedding vectors within each bin (excluding participants that emitted fewer than 10 utterances in any condition, thus ensuring balanced conditions and no leakage of utterance information across bins). This resulted in a representation of the Cinderella and Routine data as two arrays of matched size: (participants, embedding_dimension, time_bins), where embedding_dimension = 1536 and time_bins = 10. After averaging across participants within each condition, this resulted in a (number-conditions, embedding_dimension, time_bins) dimensional matrix, where the two conditions (Cinderella and Routine) are concatenated along the first dimension.

We applied dPCA on this data representation, with marginalizations separating temporal variance (shared across conditions) from condition-specific variance (Cinderella vs. Routine). dPCA regularization was determined in a data-driven fashion using the dPCA library cross-validation procedure. This decomposition yielded one set of orthogonal (decoder) principal components for temporal information, and another set of orthogonal (decoder) principal components for the condition-specific information, the latter, encompassing both a condition feature and a condition*temporal interaction feature [52]. The former maps from ambient semantic space (1536D) to a shared linear subspace that captures temporal progression independent of task condition (Cinderella or Routine).

We demonstrated the meaningfulness of these temporal subspaces in two ways. First, we quantified the amount of variance in the Cinderella/Routine embeddings that could be explained by the first (decoder) principal component of the temporal dPCA. We compared the empirically observed explained variance to a null distribution that eliminated shared temporal information between conditions. We generated this distribution through 500 permutations, in each randomly shuffling the temporal order of narratives from the Routine condition with respect to the Cinderella condition. Crucially, this approach preserves the within-condition ordering of Routine and Cinderella narratives. Statistical significance was established when the explained variance in unpermuted data exceeded the 95th percentile of the permuted distribution.

Second, we tested the ability of dPCA temporal components to generalise to unseen out-of-distribution data (a marker of external predictive validity), using 10,000 narratives from the TinyStories, as introduced above [48]. We projected each TinyStories narrative segment's semantic embedding vector onto the 1st principal component of the temporal dPCA decoding matrix, learned from the Cinderella/Routine dataset. We then trained a logistic regression model to discriminate segments in the first half of the parent narrative from the second half of the parent narrative, based on this dPCA projection alone, and quantified generalisation accuracy in 5-fold cross-validation. Statistical significance was assessed by comparing classification accuracy against a null distribution created by repeating this procedure using 500 random projections (permuting the weights of the 1st temporal dPCA axis). As a control, we repeated this procedure using the condition-specific dPCA principal axis, which is not expected to contain generalizable temporal information.

Finally, to generate an *abstraction score* for individual differences analysis, we computed the low-rank reconstruction of each narrative's semantic embedding representation, using the decoding and encoding matrices from the temporal dPCA alone (i.e., no condition-specific information). The abstraction score was computed as the amount of variance explained by the low rank temporal reconstruction ($R^2$ metric). As a control, we repeated this procedure using a condition-specific low-rank reconstruction.

## 7. LLM-derived trajectory entropy measure

For individual differences analysis, we additionally defined an LLM-based measure of narrative predictability, inspired by the framing of LLMs as "role playing agents" capable of emulating a latent data generating distribution given enough contextualising information [55–57].

For each narrative, we iteratively prompted GPT-4o-mini to generate utterance continuations conditioned on the entire narrative history up to that point. At each utterance position $t$, we provided the model with all preceding utterances (the narrative's stem) and prompted it to generate the next utterance using the following text: *"{'You are telling the story of Cinderella', 'You are describing the daily routine of an average person'}. Here is your account so far. Continue the train of thought in the same style, providing the next event in the sequence."*

For each stem, we generated 50 independent text completions (temperature = 1.2 to encourage variance, stop tokens: [".", "!", "?", ";"]). These 50 completions captured a distribution of plausible narrative continuations, considering the full narrative history. This contrasts with our alignment measure of predictability, described above, which adheres to a first-order Markovian assumption that the next point in a trajectory only depends on the location of the current point. We then applied our condition-specific schema decoders to each of the 50 generated continuations to obtain the argmax decoded event for each continuation, normalising count numbers over all continuations to generate a decoded event probability distribution.

We calculated the entropy of this event probability distribution at each stem location. High entropy indicates greater unpredictability in narrative continuation, while low entropy indicates more constrained, predictable progression. Here, predictability is computed with respect to the distribution of possible next schema events, not the specific words used to describe the event. This approach is similar to a popular *discrete semantic entropy* metric defined in the LLM interpretability literature [54].

We define the *trajectory entropy* of a given narrative as the mean of utterance-wise entropy measures, providing a single measure of trajectory predictability that accounts for the full narrative context. As expected, trajectory entropy alignment was significantly negatively correlated with trajectory alignment (correlation with schema space trajectory alignment r=-.43, p<0.001, correlation with semantic space trajectory alignment r=-.20, p<0.001, metrics averaged over conditions for each participant).

## 8. Response time (RT) analysis

For response time ($RT$) analysis, we extracted the $\log(RT)$ associated with each word, $t$. This was defined as the time delay from a space bar press following the completion of the previous word, $t-1$, and the first character of word $t$.

To assess whether inter-word RTs tracked putative utterance boundaries, we regressed the $\log(RT)$ time series onto a binary variable indicating presence of utterance boundaries in each narrative's optimal utterance parsing solution (see Utterance segmentation and vector embedding). We used linear mixed modelling to account for the nested structure of our data (individual narratives as random factor, each associated with random slopes and intercepts). We conducted this analysis separately for Cinderella and Routine conditions.

To assess whether inter-utterance RTs tracked conceptual distance (utterance separation in schema space), we used a second family of linear mixed effects models, wherein we regressed $\log(RT)$ onto a design matrix including two predictor variables: cosine distance separating consecutive utterances in schema space, and cosine distance separating consecutive utterances in semantic space (enabling an estimation in the unique variance explained by the former). This model only included timepoints associated with utterance boundaries, reflecting our focus here on inter-utterance variance in $\log(RT)$.

## 9. Individual difference analysis (Eccentricity)

For individual differences analysis, we obtained self-report psychiatric ratings from each participant (128 questions, taken from validated self-report questionnaires designed to characterise schizotypy [81], self-reported formal thought disorder [82], hallucinatory experiences [83], hypomania symptoms [84], attention deficit hyperactivity symptoms [85], obsessive compulsive symptoms [86], and general depression and anxiety [87].

As previously described [53], we used exploratory factor analysis to infer the latent transdiagnostic dimensions explaining the correlations among the 128 items, while accounting for their ordinal nature.

This approach yielded an 11-factor solution. Based on our previous work, we considered the "Eccentricity" factor to be most relevant, *a priori*, to our semantic trajectory measures. Eccentricity items pertains to a self-perceived atypicality in behaviour (*"I use long and unusual words to say simple things"*; *"People sometimes comment on my unusual mannerisms and habits"*; see [53] for full details).

To assess the predictive validity of trajectory measures, we computed the correlation coefficient between eccentricity and trajectory variables (alignment, momentum, abstraction score, discrete state sequencing). Here, as the focus was on participant-level differences, we averaged trajectory metrics across conditions for each participant. For trajectory metrics that could be defined in both schema and semantic spaces (alignment and momentum), we additionally ran multiple regression analyses including paired schema and semantic space metrics as the predictor variables. The intention here was to confirm that the predictive power of schema space metrics was preserved after accounting for variance explained by the same metric calculated in semantic space.

As control analyses, we re-computed the linear relationship between each trajectory measure and eccentricity in three further multiple regression analyses, in each controlling for a different potential confounding variable. The first potential confounding variable was *decoding* accuracy, defined as the mean number of utterances where the argmax decoder state matched the utterance's modal GPT-4o-mini label. The second variable was, *topic coverage*, defined as the degree to which a trajectory visits all schema events. Here, we calculated the cosine similarity between each trajectory point (a soft schema space embedding) and each schema event representation (a one-hot schema space embedding, with a 1 in the dimension corresponding to the schema event in question). For each schema event, we then computed the maximal cosine similarity over all trajectory points (i.e., the region of maximal trajectory proximity to a given event). Topic coverage was defined as the mean of these maximal cosine similarity scores over all events. The third variable was *utterance parsing quality*, operationalized as the degree to which detected utterance boundaries predict RT slowing (i.e., the trajectory-wise regression coefficient relating utterance boundary location to RT slowing).

# 10. Alternative approaches to concept decoding

The Concept Decoding approaches showcased in our primary results use various supervised dimensionality approaches, such as classification and dPCA, to yield cognitively aligned schema spaces. In supplementary analyses, we explored the ability of alternative approaches to uncover behaviourally relevant representations from LLM embeddings, namely BERTopic-based unsupervised topic modelling and prompt-based utterance contextualization. While these show considerable promise, our supplementary analyses reveal they are brittle to hyperparameter changes in ways that are difficult to predict *a priori*, limiting their current practical application.

### BERTopic: identifying conceptual structure using unsupervised clustering

The first approach uses unsupervised topic modelling to identify semantic clusters. These clusters – often interpreted as topics - can be thought of as proxies for schema events used in Concept Decoding (the latter, identified using a LLM-guided procedure, see [Concept decoding](#)).

We used BERTopic, a state-of-the-art unsupervised topic modelling method that identifies latent topic structure by applying clustering algorithms to LLM-derived semantic embeddings [42]. BERTopic is a highly flexible, stochastic pipeline, with many tunable hyperparameters. We used a standard implementation that begins with nonlinear dimensionality reduction on utterance-wise semantic space embeddings using UMAP [88], followed by HDBSCAN-based hierarchical clustering. This assigns each semantic embedding vector (utterance) to one topic cluster (or a single "noise" cluster). The total number of clusters is determined by UMAP and HDBSCAN hyperparameters, and subtle changes in these hyperparameters can yield vastly different topic solutions [43]. To account for this sensitivity, we implemented a grid search over BERTopic's key parameters: UMAP nearest neighbors = $25 \cdot 2^n$, where $n = [0, 1, 2, 3, 4, 5, 6, 7, 8]$; and HDBSCAN minimum cluster size = $25 \cdot 2^m$, where $m = [0, 1, 2, 3, 4, 5, 6, 7, 8]$ (HDBSCAN minimum samples are set equal to minimum cluster size). Other BERTopic parameters were fixed at common default values (UMAP number of retained dimensions = 5, UMAP ambient distance metric = "cosine", UMAP effective minimum distance between embedded points = 0.1, HDBSCAN metric = "euclidean"). For each hyperparameter setting, we ran the BERTopic algorithm over the concatenated semantic embeddings of all trajectories within a condition (semantic space embeddings from OpenAI's *text-embeddings-3-small*).

This procedure yielded 45 topic solutions, one for each hyperparameter setting, for each task condition. Each topic solution yielded a different number of semantic clusters (topics) per task condition and assigned each utterance to one cluster using both hard (argmax) and soft (probability distribution) labels (Cinderella: number topics ranged from 2-114 (median utterances per topic across BERTopic solutions ranged from 84 to 12838); Routine: number of topics = 2-86 (median utterances per topic across BERTopic solutions ranged from 81 to 11263)). Each topic solution can be thought of as a hypothesis about the conceptual structure governing the task, grouping utterances into narrative concepts (akin to schema events) at different levels of resolution (Figure S5A).

Our key question was whether BERTopic could uncover topic solutions that approximated the predictive validity of our Concept Decoding approach. We assessed this using a metric we term *transition typicality*, which can be thought of as a variant of the discrete state sequencing measure (defined above) that does not assume a ground truth state ordering (which cannot be gleaned from BERTopic topic solutions). For a given BERTopic hyperparameter setting, we represented each narrative as a (n_utterances, n_topics) matrix of soft topic assignments (yielded by the hdbscan *all_points_membership_vectors()* function), and computed a (topic, topic) joint probability matrix (as in Discrete state sequencing). We then quantified the degree to which a given trajectory's transition dynamics aligns with the group average, defining transition typicality as the Pearson's correlation between a given narrative's (topic, topic) joint probability matrix, and the mean matrix derived from all other trajectories in the sample (excluding self-transitions from this computation) (Figure S5B).

We used transition typicality to gauge BERTopic solution quality by computing the main effect of transition typicality for the solution as a whole: the t-statistic of a one-sample t-test over all trajectories, quantifying the strength of the effect that transition typicality exceeds 0 in each BERTopic solution. This can be thought of as a measure of the ability of a BERTopic solution to uncover a cognitively meaningful latent task structure. Strikingly, we found that transition typicality scores were numerically higher when computed using the VECTOR topic solution (where topics correspond to schema events yielded by our Concept Decoding step), compared to any of the 45 BERTopic topic solution (Cinderella mean transition typicality scores = 0.70 (VECTOR) vs. 0.16 - 0.66 [BERTopic]; Routine mean transition typicality scores = 0.52 (VECTOR) vs. 0.14 - 0.47 [BERTopic]) (Figure S5C).

Importantly, this predictive capacity scaled with the degree to which the BERTopic solution overlapped with the VECTOR-based topic solution. We quantified this overlap as the Normalized Mutual Information (NMI) between BERTopic and Concept Decoding cluster assignments for each utterance (the former using the one-hot BERTopic topic labels, the latter using the modal GPT-4o-mini labels) ([Figure S5](#)D, we obtain similar results using an alternative measure of topic solution overlap: the Adjusted Rand Index.).

We also quantified the degree to which individual differences in transition typicality predict eccentricity scores, again finding that this complementary measure of predictive validity also correlated with the degree to which BERTopic solutions overlapped with VECTOR-based solutions ([Figure S5](#)D).

## Contextualizing utterances: augmenting semantic embeddings using text prompts

The second approach is to inject implicit contextualising information into the utterance text itself, prior to sentence embedding. We generated context-augmented semantic embeddings by prepending each utterance with a condition-specific contextualising prompt: *"In the story of Cinderella: "* and *"In the daily routine of an average person: "*. Thus, the utterance *"The clock struck midnight"*, becomes *"In the story of Cinderella: the clock struck midnight"*. We then generated a new 1536D semantic space embedding vector for each utterance using OpenAI's *text-embeddings-3-small* model.

As expected, introducing more context into each utterance's semantic embedding vector warped the representational geometry of the embedding space, markedly improving the trajectory alignment of Cinderella trajectories in particular ([Figure S6](#)A).

In quantitative testing, we found that the context-augmented embeddings displayed a representational geometry that was closer to that of the schema space yielded by VECTOR's Concept Decoding step, compared to non-contextualised (standard) semantic embeddings ([Figure S6](#)B). We quantified this *representational* alignment for each trajectory separately, by first computing the (utterance, utterance) cosine distance in schema and (context-augmented) semantic space, and then calculating the spearman correlation between the upper triangles of these paired distance matrices. Individual-trajectory correlation coefficients were r-to-z transformed prior to applying statistical tests for differences in mean schema-space alignment between context-augmented vs. standard semantic space embeddings.

Context-augmented semantic embeddings also showed greater *trajectory* alignment and momentum compared to their non-contextualised counterparts (as defined in [Measures of trajectory organisation](#)). Nevertheless, these measures of representational quality were still significantly lower than those yielded by schema space embeddings ([Figure S6](#)C-D). Moreover, schema space embeddings showed superiority in predicting eccentricity scores (individual differences) in multiple regression analyses that included both schema metrics and context-augmented semantic embedding metrics as predictors ([Figure S6](#)E).

Contextualizing utterance embeddings in this manner comes with its own set of implicit hyperparameters: namely, the precise sequence of tokens used as the contextualising prompt. Indeed, in our own testing we found that seemingly innocuous changes to the contextualising prompt could degrade representational alignment to schema space geometry ([Figure S6](Figure S6)E). To the best of our knowledge there are no methods that enable us to predict *a priori* which prompts will yield better representational spaces, and it is beyond the scope of this paper to perform anything approximating a search over prompt space to identify optimal contextualising prompts.

## 11. General statistical analysis

All analysis was conducted in Python (v3.10.10) using standard scientific computing packages (sklearn v1.2.2, scipy v1.10.1, numpy v.23.5, umap-learn 0.5.4, hdbscan 0.8.29). For dPCA we used the method described in [52]: https://github.com/machenslab/dPCA. Where a method is not described in detail in our manuscript or referenced to prior work, the reader is to infer that we used an in-built function from these libraries (e.g., correlations, normalised mutual information, entropy, regression, PCA).

We accessed the open-source bert-based-uncased LLM through the Python Huggingface library (used for stop token probability estimation in utterance parsing). All other LLM-based analyses were conducted using OpenAI's pre-trained foundation LLMs (GPT-4o-mini-2024-07-18) and semantic embedding model (text-embeddings-3-small), accessed using the Python openai library (v1.30.1).

Throughout, we define statistical significance at an alpha=0.05, two-tailed threshold, unless otherwise stated.

We assessed main effects and group differences in central tendency using parametric *t*-tests when data met normality assumptions (as determined by the Shapiro–Wilk test). For non-normally distributed data, we used the Wilcoxon signed-rank test for one-sample or paired comparisons, and the Mann–Whitney U test for independent (unpaired) comparisons.

For analyses examining individual difference correlations with eccentricity, we used Pearson correlation coefficients when both variables were normally distributed, and Spearman rank correlations when normality assumptions were violated, as determined by the Shapiro–Wilk test.

Where we use permutation tests for significance testing, these are described in the relevant sections.

## 12. Data visualization

When visualizing the effect of utterance boundaries on inter-word RTs ([Figure 1](Figure 1)B), we linearly interpolated the sequence of RTs within each utterance to a length-10 vector to account for different number of words in each utterance (a procedure used for illustration purposes only; see main text for formal statistical analysis).

For the main trajectory plots (e.g., [Figure 2](Figure 2)), we use a convention of plotting data on the 1$^{st}$ 2 principal components of a PCA analysis on the (utterance, embedding_dimension) matrix (as described above). We highlighted exemplar narratives in a data-driven fashion. In [Figure 2](Figure 2)A, we selected the top 10 narratives in each space exhibiting minimal curvature and traversing ≥75% schema topics. In [Figure](Figure) 2C and Figure 2D, we selected exemplar trajectories corresponding to the top (black) and bottom-3 (red) narratives according to each metric (alignment, momentum).

Flow fields in Figure 2 were inspired by [80]. We divided 2D representational space into an evenly spaced grid (20 x 20) and defined 400 spatial states centred on each grid intersection (as for alignment metric definition). For each state we defined the set of all trajectory edges passing through the state. The state's mean alignment direction was the average of all these intersecting edge vectors (each edge expressed as a unit displacement vector). We displayed this information as a flow field: an arrow anchored to each state's centroid, where arrow length is proportional to the mean alignment * number of intersecting edges, and the colour intensity reflects mean alignment value alone.

When plotting the topic locations in representational space (Figure 2A), we computed each topic's centroid as the mean embedding vector of all utterances assigned to each topic by schema decoding. Radial histograms display the spatial displacement distribution of trajectory segments starting in the vicinity of each topic centroid (specifically, where the first utterance of the segment lies in a circle centred on the topic centroid, with a radius 0.1* the spatial extent of the space).

## 13. Data and code availability

Data and code to recreate paper analyses and figures will be uploaded to a public repository at the time of publication. Data will include utterance-wise semantic embedding vectors, LLM-generated labels, and RT measures. Per study protocol and the consent form, participant's narratives are available upon request from Isaac Fradkin.